\documentclass[11pt]{article}

\usepackage[titletoc,toc]{appendix}

\newcommand{\be}{\begin{equation}}
\newcommand{\ee}{\end{equation}}
\usepackage{amsmath}
\usepackage{amssymb}
\usepackage{latexsym}
\usepackage{epsfig}
\usepackage{color}
\usepackage[makeroom]{cancel}
\usepackage{breqn}
\usepackage{mathtools}
\usepackage{cite}

\newcommand{\M}{\mathcal{M}}

\newcommand{\bkt}[1]{\left(#1\right)}

\newcommand{\RRR}{{\hbox{\rm R\kern-2.35mm R}}}
\newcommand{\p}{\partial}

\def\ZZZ{{\hbox{ Z\kern-1.6mm Z}}}

\newcommand{\nin}[1] {\underline{\phantom{h}}\hskip-6pt {#1}}

\newcommand{\sectiono}[1]{\section{#1}\setcounter{equation}{0}}

\def\E{{\cal E}}      \def\H{{\cal H}}   
      \def\M{{\cal M}}

\def\?{\varphi}  \def\è{\varpi}	\def\?{\vartheta}

\makeatletter
\def\munderbar#1{\underline{\sbox\tw@{$#1$}\dp\tw@\z@\box\tw@}}
\makeatother

\begin{document}

\begin{titlepage}
\rightline{July 2016} 
\rightline{\tt MIT-CTP-4815} 
\begin{center}
\vskip 2.5cm

{\Large \bf {On the curious spectrum of \\[1ex]
duality invariant higher-derivative gravity }}

 \vskip 0.5cm

  \vskip 2.0cm
 {\large {Olaf Hohm,$^1$ Usman Naseer,$^2$ and Barton Zwiebach$^2$}}
 \vskip 1cm

{\em $^1$ \hskip -.1truecm Simons Center for Geometry and Physics, \\
Stony Brook University, \\
Stony Brook, NY 11794-3636, USA \vskip 5pt }

{\em $^2$ \hskip -.1truecm Center for Theoretical Physics, \\
Massachusetts Institute of Technology\\
Cambridge, MA 02139, USA \vskip 5pt }

ohohm@scgp.stonybrook.edu,   \  unaseer@mit.edu, \  zwiebach@mit.edu\\

\vskip 2.5cm
{\bf Abstract}

\end{center}

\vskip 0.5cm

\noindent
\begin{narrower}

\baselineskip15pt

We analyze the spectrum of the exactly duality and gauge invariant 
higher-derivative double field theory.  
While this theory is based on a chiral CFT and 
does not correspond to a standard string theory,  
our analysis illuminates a number of issues central in string theory.   
The full quadratic action is rewritten as a two-derivative theory with additional fields. 
This allows for a simple  analysis of the spectrum, which contains 
two massive spin-2 ghosts and massive scalars, in addition to the massless fields. 
Moreover, in this formulation, the massless or 
tensionless limit $\alpha'\rightarrow \infty$ 
is non-singular and leads to an enhanced gauge symmetry.  
We  show that the massive modes can be   
integrated out exactly at the quadratic level, 
leading to an infinite series of higher-derivative corrections. 
Finally, we present a ghost-free massive extension 
of linearized double field theory, 
which employs a novel mass term for the dilaton and metric.

\end{narrower}

\end{titlepage}

\baselineskip=16pt
\parskip=\medskipamount

\setcounter{tocdepth}{2}
\tableofcontents

\baselineskip15pt


\vspace{2ex} 

\sectiono{Introduction}

Some of the salient characteristics of string theory   
are the presence of higher-derivative 
$\alpha'$ corrections, massive modes of higher spin,    
and duality invariance, such as T-duality. 
In this paper we aim to illuminate the interplay between these aspects by analyzing the quadratic approximation 
to the $\alpha'$-deformed double field theory (DFT) constructed by Siegel and two of the authors in \cite{Hohm:2013jaa} 
(generalizing \cite{Siegel:1993th,Hull:2009mi,Hohm:2010jy,Hohm:2010pp,Hohm:2010xe} and further investigated in \cite{Hohm:2014eba,Hohm:2014xsa,Marques:2015vua,Hohm:2015mka,Hohm:2015doa,Siegel:2015axg,Naseer:2016izx,Huang:2016bdd,Casali:2016atr}). 
This theory, henceforth called HSZ theory,  
 contains higher-derivative corrections and is exactly duality\footnote{In the following we 
refer to the global $O(d,d,\mathbb{R})$ invariance emerging upon dimensional reduction on a torus, which is 
made manifest in DFT, for brevity simply as `duality invariance'.} and gauge invariant 
and hence well-suited for this purpose. 

HSZ theory 
is written in terms of a duality-invariant dilaton $\phi$ and an unconstrained
 `double metric' ${\cal M}_{MN}$, with $O(D,D)$ indices $M,N=1,\ldots, 2D$, which in turn can be decomposed  
 into a 
 generalized metric ${\cal H}_{MN}$, taking values in $O(D,D)$ and 
 encoding the spacetime metric and two-form field, 
 {\em plus}   additional fields.  
 Specifically, expanding around flat space and reducing to $D$-dimensional indices $i,j,\ldots$, 
 the metric and two-form fluctuations are encoded in a general second-rank tensor $e_{ij}$, while the additional fields  
 are given by two symmetric tensors $a_{ij}$ and $\bar{a}_{ij}$. 
 
 As was outlined in \cite{Hohm:2013jaa} and shown in more detail in \cite{Hohm:2015mka}, these extra fields can be 
 treated as auxiliary fields in that they can be eliminated algebraically by iteratively solving their field 
 equations in terms of the massless fields. This leads to an infinite number of higher-derivative $\alpha'$ corrections 
 for the massless fields. To be more specific, consider the Lagrangian, which to  lowest order in derivatives contains 
 the potential terms 
\be\label{potentiaL}
L \ = \ \tfrac{1}{2\alpha'}\, e^{\phi} \bkt{  a_{ij} a^{ij} -\bar{a}_{ij} \bar{a}^{ij} 
\ + \ \bar{a}^{ij} e_{ki}e^k{}_{j}
\ - \ a^{ij} e_{ik}e_{j}{}^k\ +\ \tfrac{1}{3} a^{ij} a_{i}{}^k a_{jk} \ - \ 
\tfrac{1}{3} \bar{a}^{ij} \bar{a}_{i}{}^k \bar{a}_{jk}} \ + \ \cdots  \;, 
\ee
where the dots represent terms that contain from two to six derivatives. 
The field equations for 
$a$ and $\bar{a}$, to lowest order in $\alpha'$  and to lowest order in the number of massless fields,  
imply $a_{ij}=\frac{1}{2}e_{ik}e_{j}{}^k$ 
and $\bar{a}_{ij}=\tfrac{1}{2}e_{ki}e^k{}_{j}$, 
so that they may be eliminated in terms of the massless fields. In principle, this procedure can be extended 
to any order in $\alpha'$ and any number of (massless) fields. 

It is not yet known how the complete HSZ  
theory looks  
in terms of conventional 
gravity fields and after eliminating the extra fields, but it was shown that to first order in 
$\alpha'$ it encodes 
the gravitational Chern-Simons modification implied by the Green-Schwarz mechanism \cite{Hohm:2014eba,Hohm:2014xsa} 
and to second order in $\alpha'$ it contains  
a cube of the Riemann tensor \cite{Naseer:2016izx}. 
As such, the theory encodes elements both of heterotic string theory (the Green-Schwarz deformation) 
and of bosonic string theory (the Riemann-cube term), while lacking structures present in both 
(like a Riemann-squared term). Therefore, this theory does not correspond to any conventional string theory, 
in line with the non-standard chiral CFT on which it was based. We expect, however, that a more general class 
of gauge and duality invariant theories exists for which the conventional string theories would 
arise as particular combinations, as shown to first-order in $\alpha'$ at the cubic level in \cite{Hohm:2014xsa}
and more recently to all orders in fields in \cite{Marques:2015vua}.

The goal of the present paper is to analyze the dynamical content (the particle spectrum) of 
HSZ  
theory, 
including higher derivatives but restricting to the quadratic approximation around flat space. 
It was recently pointed out in  \cite{Huang:2016bdd} that keeping the extra fields $a_{ij}$ and $\bar{a}_{ij}$, rather than integrating them out, 
indicates  
the presence of 
two massive spin-2 modes in the spectrum. Indeed, one reads off quadratic mass terms from the 
Lagrangian (\ref{potentiaL}), while the `higher-derivative' terms starting with two derivatives naturally 
yield the kinetic terms. 
Qualitatively, this seems to match the spectrum of the `chiral string theory' 
investigated in \cite{Huang:2016bdd}, whose spectrum also contains two massive spin-2 modes. 
In this paper we will compute the spectrum of the 
exact 
quadratic theory, which includes up to 
six derivatives, and confirm the presence of massive spin-2 states. However, the detailed spectrum differs 
from that of the chiral string theory given in \cite{Huang:2016bdd}.

Our analysis is simplified by introducing 
further fields that allow us to reduce the number of derivatives to two. 
In order to elucidate the 
structure of these theories, we find it convenient to compare them with 
a massive deformation of the original (massless) 
linearized DFT. This theory, which seems 
interesting in its own right, is given by the Lagrangian 
\be\label{massiveDFTINTRO}  
  L_{\text{mDFT}} \ = \  \tfrac{1}{2}e^{ij}{\cal R}_{ij}(e,\phi)+\tfrac{1}{2}\phi{\cal R}(e,\phi) 
  -\tfrac{1}{4}M^2(e^{ij} e_{ij}-  4 \phi^2)\;, 
 \ee       
where ${\cal R}_{ij}$ and ${\cal R}$ are the linearized Ricci tensor and scalar curvature of DFT, 
whose explicit forms are given in (\ref{eq:LinScalarR}).  
We will show that this model propagates precisely a massive 
spin-2 mode, a massive
two-form 
field, and a massive scalar, without any undesired or ghost-like modes. 
This result hinges on the structure of both the mass terms and the kinetic terms, 
which are such that in the massless limit $M^2\rightarrow 0$ the theory is invariant under the 
DFT gauge symmetry, 
 \be\label{DFTmassless}
  \delta e_{ij} \ = \ D_i\bar{\lambda}_j+\bar{D}_{j}\lambda_i\;, \qquad 
    \delta \phi \ = \ \tfrac{1}{2} (D_i\lambda^i+\bar{D}_i\bar{\lambda}^i)\;. 
      \ee  
Intriguingly, this model seems new as it is \textit{not} field-redefinition
equivalent to  the Fierz-Pauli-theory of 
 (linearized) massive gravity augmented by a massive two-form and a massive scalar.
Indeed, while 
the kinetic terms in (\ref{massiveDFTINTRO}) can be diagonalized 
(returning to Einstein frame) in order to 
write the model as a sum 
of linearized gravity, massless two-form and massless scalar, one cannot simultaneously 
diagonalize the above mass terms.
Nevertheless, the above model is ghost-free, and this may shed a new light 
on the old problem of finding a consistent non-linear theory of massive gravity (see \cite{deRham:2014zqa}
for a recent review). 

Remarkably, 
the six-derivative HSZ quadratic Lagrangian 
can be rewritten as 
a two-derivative Lagrangian by 
introducing two auxiliary scalars $\varphi$ and 
$\bar{\varphi}$, which pair up with $a_{ij}$ and $\bar{a}_{ij}$, to play a 
role largely analogous to that 
which 
the dilaton $\phi$ plays for 
$e_{ij}$. 
In particular,  
thanks to these new fields, 
the kinetic terms are `improved' relative to the original 
two-derivative terms  
and  
the number of degrees of freedom does not increase. The massive spin-2 modes are ghost-like, as can be seen from the  
overall sign of the kinetic terms. The presence of ghost-like massive spin-2 modes is in qualitative agreement 
with the chiral string theory \cite{Huang:2016bdd} but, again, the detailed spectrum differs.

The improved structure of the kinetic terms is reflected by an enhanced gauge invariance in the 
massless limit, as for the massive DFT theory  
above. This symmetry reads 
 \be\label{gaugeNEWINTRO}
 \begin{split}
  \delta_{\zeta}a_{ij} \ &= \ D_{i}\zeta_j+D_j\zeta_i\;, \qquad
  \delta_{\zeta}\varphi \ = \  -D_i\zeta^i\;, \\[0.5ex]
  \delta_{\bar{\zeta}}\bar{a}_{ij} \ &= \ \bar{D}_{i}\bar{\zeta}_j+\bar{D}_j\bar{\zeta}_i\;, \qquad
  \delta_{\bar{\zeta}}\bar{\varphi} \ = \  \ \bar{D}_i\bar{\zeta}^i \;, 
 \end{split}
 \ee 
and thus takes the form of two additional diffeomorphism-like symmetries with parameter $\zeta_i$
and $\bar{\zeta}_i$.  Note that the massless limit corresponds to the tensionless limit 
$\alpha'\rightarrow \infty$ and hence this model confirms the general expectation that string theory 
exhibits an enlarged gauge symmetry in this limit \cite{Gross:1988ue}.

We close this introduction 
with some general remarks. 
Given the presence of ghost-like modes in the spectrum, 
it follows that this theory is problematic  
--- at least around 
flat space and to the extent that the quadratic theory provides a reliable approximation. 
It should be recalled, however, that 
the inclusion of more than two derivatives generically leads to additional 
propagating degrees of freedom, which are typically ghost-like and massive. For instance, 
the addition of curvature-squared terms to the Einstein-Hilbert action generally 
leads to a massive spin-2 ghost and a massive scalar, thereby violating unitarity.
Can the spin-2 ghosts in HSZ theory be interpreted similarly?   
We will show in sec.~6 that in the quadratic theory the massive fields can be integrated out exactly. 
Due to the presence of \textit{two} massive 
spin-2 fields 
this leads to 
an \textit{infinite} number of higher-derivative corrections.

In the usual string field 
theories one can always choose a field basis for which the 
propagator is not modified, 
making manifest  
that there is no conflict with unitarity. 
To first order in $\alpha'$, 
one employs  
 the Gauss-Bonnet combination~\cite{Zwiebach:1985uq}, which is a total derivative at the quadratic 
level.\footnote{Other higher-derivative theories that do not propagate ghosts are 
Einstein-Hilbert plus 
the square of the pure Ricci scalar, which is equivalent to a massive scalar coupled to gravity 
and currently a favored model for inflation (Starobinsky model) \cite{Planck:2013jfk}, 
and new massive gravity in $2+1$ dimensions, 
which augments a `wrong-sign' Einstein-Hilbert term with a particular 
curvature-squared term \cite{Bergshoeff:2009hq}.} 
In contrast, 
there is evidence that any theory 
that is not a complete string theory (like generic higher-derivative gravity) 
is problematic   
at some level, see e.g.~\cite{Camanho:2014apa}. Our findings here seem to confirm this.

It would be instructive to  
investigate the physical content around other, 
curved backgrounds. It may well   
be that some form of `ghost condensation' takes place, so that the flat-space ghosts  
disappear on other backgrounds \cite{ArkaniHamed:2003uy}. A 
simple version 
of such a phenomenon is already 
visible for the flat space theories analyzed here: 
the potential (\ref{potentiaL}) allows for two different flat space solutions, corresponding 
to sending the background generalized metric $\bar{\cal H}$ to $-\bar{\cal H}$
 (see section~\ref{fullndlagvac}).  
Intriguingly, ghosts in one vacuum become healthy in the other and vice versa.

\sectiono{Full quadratic theory and non-derivative terms }\label{fqtandt}

In this section we compute the full quadratic Lagrangian and the potential of 
HSZ theory~\cite{Hohm:2013jaa}.    
From the quadratic Lagrangian we will see that the theory has both `ghost-like' and `healthy' degrees of freedom. By analyzing the potential, we  show that the theory admits two vacua 
with constant backgrounds. 
Both of these vacua have the same number of degrees of freedom;  `ghost-like' fields of one vacuum, however, correspond  
 to `healthy' fields of the other vacuum and vice versa.

The full action for the HSZ theory can be written as \cite{Hohm:2013jaa}
\be\label{eq:ActionFull}
S\ =\ \int e^{\phi} L\;,  \ \ \ \ \ \ \ L =\tfrac{1}{2} \text{tr}\bkt{\mathcal{T}}\ -\tfrac{1}{6}\langle \mathcal{T}|\mathcal{T}\star \mathcal{T}\rangle\;,
\ee
where $\mathcal{T}$ is a tensor operator which encodes the double metric $\mathcal{M}$. The Lagrangian $L$ has terms with up to six derivatives, given in equation (3.16) and (3.17) of \cite{Naseer:2016izx}. 
This Lagrangian can be expanded around a {\it constant} background  $\langle \mathcal{M}\rangle$ that can be identified with a constant generalized metric \cite{Hohm:2014xsa}:
\be 
\M_{MN}\ = \ \bar{\H}_{MN}\ + m_{MN}\ =\ \ \bar{\H}_{MN}\ + m_{\nin{M}\bar{N}}\ + m_{\nin{N}\bar{M}}\ + a_{\bar{M}\bar{N}}\ + a_{\nin{M}\nin{N}}\;.\label{eq:doubleMetricExp}
\ee
Here $\bar{\H}$ is the constant background generalized metric, and $m_{MN}$ encodes the double metric fluctuations around this background. These fluctuations have been decomposed using projected $O(D,D)$ indices defined as 
\be\label{eq:ProjectedInd} 
V_{\nin{M}} \ = \ P_{M}{}^{N} V_N\,,\ \ \ \ \ V_{\bar{M}}\ = \ \bar{P}_{M}{}^{N} V_{N}\; ,
\ee
and analogously for higher tensors, with the projectors 
\be\label{eq:ProjectorDef}
P_M{}^N \ = \ \tfrac{1}{2} \bkt{\eta-\bar{\H}}_{M}{}^N\;,\ \ \ \ \ \bar{P}_M{}^N\ = \ \tfrac{1}{2}\bkt{\eta+ \bar{\H}}_M{}^N\;.
\ee

Fluctuations of the double metric can be related to conventional fields with spacetime indices as explained in detail in section 5.3 of \cite{Hohm:2014xsa}. Based on equation (5.57) of \cite{Hohm:2014xsa}, we introduce the `conventional' counterparts for fields $a_{\nin{M}\nin{N}}$ and $a_{\bar{M}\bar{N}}$  as follows:
\be\label{eq:aabardef}
a_{ab}\ =\ \tfrac{1}{2} \E_a{}^M \E_b{}^N a_{\nin{M}\nin{N}}\;, \ \ \ \ \ \ \bar{a}_{\bar{a}\bar{b}}\ =\ \tfrac{1}{2} \E_{\bar{a}}{}^M \E_{\bar{b}}{}^N a_{\bar{M}\bar{N}}\;.
\ee
Here $\E_{A}{}^M= \bkt{\E_a{}^M, \E_{\bar{a}}{}^M}$ is the background vielbein and $\bkt{a,\bar{a}}$ are flat frame indices. The particular choice of the background vielbein made in equation (5.42) of \cite{Hohm:2014xsa} allows one to identify the flat and curved indices. The rules for translating an expression written in terms of projected indices to an expression written in terms of conventional spacetime indices can be summarized as follows:
\begin{itemize}
\item Replace $m_{\nin{M}\bar{N}}$ by $e_{mn}$, $a_{\nin{M}\nin{N}} $ by $a_{mn}$ and $a_{\bar{M}\bar{N}}$ by $\bar{a}_{mn}$.
\item Replace under-barred derivatives by $D$ and barred derivatives by $\bar{D}$ defined as in \cite{Hohm:2014xsa}, 
\be
D_i\ =\ \p_i - E_{ik} \tilde{\p}^k\; , \qquad \bar{D}_i\ =\ \p_i \ + E_{ki} \tilde{\p}^k\ , 
\ee
where  $E_{ij}=  G_{ij}\ + B_{ij}$ is given in terms of the 
constant background metric and $b$-field. 
The strong constraint takes the form $D^i D_i = \bar{D}^i \bar{D}_i$, 
acting on arbitrary fields and all their products. 
\item Multiply by a coefficient, which is the product of a factor of $2$ for each $m$, $a$, or $\bar{a}$ field, 
a factor of $+\tfrac{1}{2}$ for each barred contraction and a factor of $-\tfrac{1}{2}$ for each under-barred contraction.
\end{itemize}

\subsection{Full quadratic Lagrangian}   

The zero- and two-derivative 
parts of the HSZ quadratic action have been computed previously in \cite{Hohm:2014xsa} (see 
eq.~(5.7)) and in \cite{Naseer:2016izx} (see eqs.~(4.7) and (4.10)). In terms of conventional fields, they are 
\be\label{eq:L20L22}
\begin{split}
L^{( 2, 0)}\  &= \  \tfrac{1}{2}\,  a^{ij} a_{ij}\ -\tfrac{1}{2}\,  \bar{a}^{ij}\bar{a}_{ij}\; , 
\\
L^{( 2, 2) } \  &= \ \tfrac{1}{4} e^{ij} \square e_{ij}
+\tfrac{1}{4} \bkt{D_i e^{ij}}^2 
+\tfrac{1}{4}\bkt{\bar{D}_j e^{ij}}^2 
+e^{ij}D_i \bar{D}_j\phi -\phi\, \square \phi \\
&
\quad \; \;
- \tfrac{1}{8}  a^{ij} \, \square a_{ij} 
-\tfrac{1}{4} \bkt{D_{i}a^{ij}}^2\ 
-\tfrac{1}{8} \bar{a}^{ij}\,  \square \bar{a}_{ij}\
- \tfrac{1}{4} \bkt{\bar{D}_j \bar{a}^{ij}}^2\,, 
\end{split}
\ee
where   $\square\equiv D_i D^i = \bar{D}_i \bar{D}^i$.  From the two-derivative Lagrangian, we note that the kinetic terms for $a_{ij}$ and $\bar{a}_{ij}$ appear with the `wrong' sign and hence describe ghost-like degrees of freedom. 

The four- and six-derivative parts of the quadratic Lagrangian can be computed explicitly starting from equation (3.17) of \cite{Naseer:2016izx}.  The computation can be simplified by noting that any term which involves derivatives acting on more than two fields will not contribute to the quadratic Lagrangian. Further, terms of the form $\bkt{\M^2}_{MN}\p^M\bkt\cdots \p^N\bkt{\cdots}$ can also be ignored, because upon expanding around the background generalized metric, such a term would vanish at quadratic level due to the strong constraint.  After excluding such terms, one gets the following expressions for the four- and six-derivative terms that can contribute to the quadratic Lagrangian:
\be\label{eq:L24L26DoubleMet} 
\begin{split}
L^{(\cdot,4)}\ =\  \tfrac{1}{12}\, \mathcal{M}^{M N}  \Bigl[& {\partial}_{M}{\mathcal{M}^{P Q}}\,  {\partial}_{P Q}\,^{K}{\mathcal{M}_{N K}}\,  
 + 2\, {\partial}^{P}{\mathcal{M}_{M}\,^{Q}}\,  {\partial}_{N Q}\,^{K}{\mathcal{M}_{P K}}\, 
- \, {\partial}_{M}{\mathcal{M}^{P Q}}\,  {\partial}_{N P}\,^{K}{\mathcal{M}_{Q K}}\, 
\\[1ex] 
 &+{\partial}_{M N}{\mathcal{M}^{P Q}}\,  {\partial}_{P}\,^{K}{\mathcal{M}_{Q K}}\,
-  \, {\partial}_{M}\,^{P}{\mathcal{M}^{Q K}}\,  {\partial}_{N Q}{\mathcal{M}_{P K}}\, 
- \,  {\partial}_{M}\,^{P}{\mathcal{M}_{P}\,^{Q}}\,  {\partial}_{N}\,^{K}{\mathcal{M}_{Q K}}\, 
\\[1ex] 
&+ \mathcal{M}^{P Q} \Bigl({\partial}_{M}{\mathcal{M}_{P}\,^{K}}\,  {\partial}_{N Q K}{\phi}\,  
 - 2\, {\partial}_{M}\,^{K}{\mathcal{M}_{P K}}\,  {\partial}_{N Q}{\phi}\,  
+3\,{\partial}_{M N}{\mathcal{M}_{P}\,^{K}}\,  {\partial}_{Q K}{\phi}\,
\\[1.0ex]
&\qquad\qquad + 3\,{\partial}_{N}{\mathcal{M}_{M}\,^{K}}\,  {\partial}_{P Q K}{\phi}\, 
+ 3\,  {\partial}^{K}{\mathcal{M}_{M K}}\,  {\partial}_{N P Q}{\phi}\Bigr)\Bigr]\, + \cdots , \\[2ex]
L^{(\cdot,6)}\  = \ \tfrac{1}{48}\, \mathcal{M}^{M N} \big(&{\partial}_{M}\,^{P Q}{\mathcal{M}^{K L}}\,  {\partial}_{N K L}{\mathcal{M}_{P Q}}\,  
+6 \,{\partial}_{M}\,^{P}{\mathcal{M}^{Q K}}\,  {\partial}_{N Q K}\,^{L}{\mathcal{M}_{P L}}\,  
- 2 {\partial}^{P Q}{\mathcal{M}^{K L}}\,  {\partial}_{M N K L}{\mathcal{M}_{P Q}}\big)
  \\[1ex] &\hspace{-2cm}
  + \tfrac{1}{8}\, \mathcal{M}^{M N} \mathcal{M}^{P Q} \bkt{{\partial}_{M P}{\mathcal{M}^{K L}}\,  {\partial}_{N Q K L}{\phi}\,
  -\,{\partial}_{M K L}{\mathcal{M}^{K L}}\,  {\partial}_{N P Q}{\phi}\,  
-2\,  \mathcal{M}^{K L} {\partial}_{M N P}{\phi}\,  {\partial}_{Q K L}{\phi}\,} + \cdots,
\end{split}
\ee
where `$\cdots$' denotes terms which do  
 not contribute to the quadratic Lagrangian  and
 $\p_{M_1M_2\cdots M_k} \equiv \p_{M_1}\p_{M_2}\cdots \p_{M_k}$. 
In computing this Lagrangian from (3.17) in~\cite{Naseer:2016izx}, 
no integrations by part have been performed.
After expanding around the background generalized metric and 
keeping only 
terms quadratic in fields, we get:
\be \label{eq:L2426Expanded}
\begin{split}
L^{(2,4)} \ &= \ 
-\tfrac{1}{4}\, a^{\bar{M}\bar{N}}\p_{\bar{M}\bar{N}\bar{P}{\bar{Q}}}a^{\bar{P}\bar{Q}}\ 
+ \tfrac{1}{4}\,  a^{\nin{M}\nin{N}}\p_{\nin{M}\nin{N}\nin{P}\nin{Q}} a^{\nin{P}\nin{Q}}\ 
+ \tfrac{1}{4}{\cal R}\ \p_{\bar{P}\bar{Q}}a^{\bar{P}\bar{Q}}\ 
-\tfrac{1}{4}{\cal R}\ \p_{\nin{P}\nin{Q}}a^{\nin{P}\nin{Q}}\;, \\
L^{(2,6)} \ &= \ \tfrac{1}{16}\, 
\bigl(  \p_{\bar{M}\bar{N}} a^{\bar{M}\bar{N}} +\p_{\nin{M}\nin{N}} a^{\nin{M}\nin{N}} -\ {\cal R}\bigr) 
{\square} 
 \bigl(  \p_{\bar{M}\bar{N}} a^{\bar{M}\bar{N}} +\ \p_{\nin{M}\nin{N}} a^{\nin{M}\nin{N}} \  -\ {\cal R}\bigr)\;.
\end{split}
\ee
Here  ${\cal R}$ is the linearized scalar curvature, which can be written in terms of the double metric fluctuation $m_{\nin{M}\bar{N}}$ or $e_{ij}$ as follows:
\be \label{eq:LinScalarR}
\begin{split}
{\cal R} \ \equiv \ &  \ -2\p_{\nin{M}\bar{N}} m^{\nin{M}\bar{N}} -  
2{\square} 
\phi \ = \ D_i \bar{D}_j e^{ij}\ 
-
2\square \phi\,, \\[0.5ex]
{\cal R}_{ij} \ \equiv \ & \  \tfrac{1}{2} \square e_{ij}  - \tfrac{1}{2} \, D_i D^k e_{kj}
- \tfrac{1}{2} \, \bar D_j \bar D^k e_{ik}  +  D_i \bar D_j \phi \,,
\end{split}
\ee
where we included the definition of the linearized Ricci tensor for future use. These tensors  are invariant under (\ref{DFTmassless}). 
The above four- and six-derivative 
Lagrangians can be written in terms of the conventional fields and spacetime indices following the rules stated after equation (\ref{eq:aabardef}).
\be \label{eq:L2426Final}
 \begin{split}
  L^{(2,4)} 
  \ &=  \  -\tfrac{1}{16}\, \bkt{\bar{D}_i\bar{D}_j \bar{a}^{ij}}^2\ 
+ \tfrac{1}{16}\, \bkt{D_i D_j a^{ij}}^2 +\tfrac{1}{8}{\cal R}\, {\bar{D}_i\bar{D}_j \bar{a}^{ij} 
 - \tfrac{1}{8}{\cal R}\, D_i D_j a^{ij}}\;, \\[2ex] 
 L^{(2,6)} 
 \ & = \ 
 \tfrac{1}{64}\, 
 \bkt{ 
\bar{D}_i\bar{D}_j \bar{a}^{ij}\ 
+\ D_i D_j a^{ij} -2 {\cal R}}
\square \bkt{\bar{D}_i\bar{D}_j \bar{a}^{ij}\ 
+\ D_i D_j a^{ij} -2 {\cal R}}.
 \end{split}
\ee

\subsection{Full non-derivative Lagrangian and vacua}\label{fullndlagvac}

The full non-derivative part $L^{(0)}$ of the HSZ Lagrangian
is given by
\be \label{eq:L0DoubleMet}
L^{(0)}\ = \ e^{\phi}\bkt{\tfrac{1}{2} \M_M{}^M\ -\tfrac{1}{6} \M_{MN}\M^{NP} \M_{P}{}^M}.
\ee 
After expanding around the generalized metric, it can be written as:
\be \label{eq:L0Expanded}
\begin{split}
L^{(0)}\ & = \ \tfrac{1}{2}\, e^{\phi}\ \Bigl(a^{\,\nin{M}\nin{N}} a_{\nin{M}\nin{N}}\ -\,  a^{\nin{M}\nin{N}} m_{\nin{M}}{}^{\bar{P}} m_{\nin{N}\bar{P}} -\tfrac{1}{3}\, a^{\nin{M}\nin{N}} a_{\nin{M}}{}^{\nin{P}} a_{\nin{N}\nin{P}}\\ &\ \ \ \ \ \ \  
\qquad -  a^{\bar{M}\bar{N}}a_{\bar{M}\bar{N}}  -\, a^{\bar{M}\bar{N}} m^{\nin{P}}{}_{\bar{M}} m_{\nin{P}\bar{N}} \ -\tfrac{1}{3}\, a^{\bar{M}\bar{N}}a_{\bar{N}}{}^{\bar{P}} a_{\bar{N}\bar{P}}\Bigr)\;.
\end{split}
\ee
Translating  to conventional variables we get:  
\be \label{eq:L0Final}
L^{(0)}\ =\ \tfrac{1}{2} e^{\phi} \bkt{  a_{ij} a^{ij} - a^{ij} e_{ik}e_{j}{}^k\ +\tfrac{1}{3} a^{ij} a_{i}{}^k a_{jk} -\bar{a}_{ij} \bar{a}^{ij}+ \bar{a}^{ij} e_{ki}e^k{}_{j}\ -\tfrac{1}{3} \bar{a}^{ij} \bar{a}_{i}{}^k \bar{a}_{jk}}.
\ee

Let us now analyze the critical points of this potential. Specifically, we look at the critical points with 
$\langle e_{ij}\rangle  =0$, where $\langle A\rangle$ denote the value of $A$ at the critical point. The dilaton independent part of the potential has four critical points: 
\be \label{eq:criticalPts}
\begin{split}
& \langle a_{ij}\rangle\ =\ 0\;, \quad\qquad  \langle \bar{a}_{ij}\rangle  \ = \ 0\;, \\ 
&\langle a_{ij}\rangle \ =\ -2\eta_{ij} \;, \quad \langle \bar{a}_{ij}\rangle \ = \ -2\eta_{ij}\;, \\[1.0ex]  
& \langle a_{ij}\rangle \ = \ 0\; ,\quad\qquad    \langle \bar{a}_{ij} \rangle \ = \ -2\eta_{ij}  \;,\\
& \langle a_{ij}\rangle \ = \ -2\eta_{ij}\; ,\quad  \langle \bar{a}_{ij}\rangle \ = \ 0\;. 
\end{split}
\ee 

It is easy to see that the potential vanishes at the first two of these critical points and is non-vanishing at the other two. Moreover, extremizing the potential with respect to the dilaton requires the potential to be zero at the critical point. Hence, only the first two critical points correspond to true vacua. The first of these critical points 
leads to the quadratic Lagrangian discussed in the previous subsection.  

We claim that the second critical point corresponds to expanding the double metric around 
a background generalized metric with the overall sign reversed, 
$\langle \M \rangle = -\bar{\H}$. A short calculation, using equation (\ref{eq:aabardef}),   shows that  $\langle a_{ab}\rangle = -2G_{ab}$  corresponds to  $\langle a_{\nin{M}\nin{N}}\rangle = 2 P_{MN}$ and $\langle \bar{a}_{\bar{a}\bar{b}}\rangle = -2G_{\bar{a}\bar{b}}$ corresponds to $\langle a_{\bar{M}\bar{N}}\rangle = -2 \bar{P}_{MN}$. Here $G_{ab}$ is the background metric in `flattened' indices, which corresponds to $\eta_{ij}$ in curved indices. We write fields $a_{\nin{M}\nin{N}}$ and $a_{\bar{M}\bar{N}}$ as
\be \label{eq:aabarBackground}
a_{\nin{M}\nin{N}}\ =\ 2P_{MN}+a'_{\nin{M}\nin{N}}\; ,\ \ \ \  \ a_{\bar{M}\bar{N}}\ =\ -2\bar{P}_{MN}+ a'_{\bar{M}\bar{N}}\,.
\ee
Using this in the double metric expansion (\ref{eq:doubleMetricExp}) and dropping the primes, we get:
\be \label{eq:DoulbeMetricExp}
\M_{MN} \ = \ -\bar{\H}_{MN} +m_{\nin{M}\bar{N}}\ + m_{\nin{N}\bar{M}}\ + a_{\bar{M}\bar{N}}\ + a_{\nin{M}\nin{N}}\,,
\ee
proving the claim that the second critical point in equation (\ref{eq:criticalPts}) corresponds to expanding the double metric around $-\bar{\cal H}$.

The physical consequence of expanding around this critical point 
is to swap the ghost-like and healthy degrees of freedom.
To see this note that changing the sign of the background generalized metric corresponds to changing the sign of the background metric while leaving the background two-form field unchanged. Hence, the Lagrangian around this background can simply be obtained by changing the sign of the background metric used in contracting different indices, i.e.,
\be 
e^{ij} \square e_{ij} =  e_{ij}\p_k \p_l e_{mn}\ \eta^{im}\eta^{jn} \eta^{kl}\ \to -e_{ij}\p_k \p_l e_{mn}\ \eta^{im}\eta^{jn} \eta^{kl}= -e^{ij} \square e_{ij}.
\ee
Thus, 
the two derivative quadratic Lagrangian around this critical point takes the following form:
\be \label{eq:L22OtherVac}
\begin{split}
L^{(2,2)}\Big|_{\langle a\rangle=\langle \bar{a}\rangle =-2 \eta}\ &=\ -\tfrac{1}{4} e^{ij} \square e_{ij}
-\tfrac{1}{4} \bkt{D_i e^{ij}}^2 
-\tfrac{1}{4}\bkt{\bar{D}_j e^{ij}}^2 
+e^{ij}D_i \bar{D}_j\phi \\
&
\ \ \ +\phi\, \square \phi \ 
+\tfrac{1}{8} \bar{a}^{ij}\,  \square \bar{a}_{ij}\ 
+\ \tfrac{1}{8}  a^{ij} \, \square a_{ij} 
+\tfrac{1}{4} \bkt{D_{i}a^{ij}}^2\ 
+ \tfrac{1}{4} \bkt{\bar{D}_j \bar{a}^{ij}}^2. 
\end{split}
\ee
We see that the field $e_{ij}$ has kinetic terms with the `wrong' sign while those for $a_{ij}$ and $\bar{a}_{ij}$ come with the `right' sign. This is 
analogous to the phenomenon of `ghost-condensation' \cite{ArkaniHamed:2003uy}, where kinetic terms for fields have different signs in different vacua.

\sectiono{Spectrum of the quadratic theory}   

In this section we give a complete analysis of the degrees of freedom
in HSZ theory as determined by 
the full quadratic Lagrangian 
around flat space.
We begin with the two-derivative 
quadratic theory
and determine its spectrum.     Then we turn to the full six-derivative
quadratic theory and reconsider the spectrum.   The calculations
are  significantly 
simplified 
by the observation that the six derivative theory can be 
rewritten 
as a two-derivative theory with additional scalar fields.  
The analysis of the spectrum reveals that, in this case, higher
derivatives do not alter the number of degrees of freedom.
The masses of some fields, however, are changed.

\subsection{Spectrum of the two-derivative quadratic theory}

The two-derivative quadratic theory is defined by the Lagrangian in (\ref{eq:L20L22}),
where we combine all quadratic terms with two or less derivatives: 
\be
\label{bz-L22}  
\begin{split}
L^{( 2, \leq 2) } \  &=\ \tfrac{1}{4} e^{ij} \square e_{ij}
+\tfrac{1}{4} \bkt{D_i e^{ij}}^2 
+\tfrac{1}{4}\bkt{\bar{D}_j e^{ij}}^2 
+e^{ij}D_i \bar{D}_j\phi \ -\phi\, \square \phi  \\[0.5ex]
&
-\ \tfrac{1}{8}  a^{ij} \, \square a_{ij} 
-\tfrac{1}{4} \bkt{D_{i}a^{ij}}^2\  + \tfrac{1}{2\alpha'}\,  a^{ij} a_{ij}
\\[0.5ex]
&
-\ \tfrac{1}{8} \bar{a}^{ij}\,  \square \bar{a}_{ij}\,  
- \tfrac{1}{4} \bkt{\bar{D}_j \bar{a}^{ij}}^2  
 -\tfrac{1}{2\alpha'}\,  \bar{a}^{ij}\bar{a}_{ij}\;. 
\end{split}
\ee
The first line in this Lagrangian contains the familiar massless 
degrees of freedom. 
There is a massless graviton, a massless two-form field
and a massless scalar dilaton. 

On the second and third lines we have two symmetric tensors
$a_{ij}$ and $\bar a_{ij}$ with mass terms.   This quadratic
two-derivative action does not match the Fierz-Pauli Lagrangian
by a long shot.  In that theory the non-derivative terms are 
those of a massless spin two field, and we do not have those terms.
Moreover, the two-derivative terms have the wrong sign,
as can be seen comparing with those for $e_{ij}$. 
The Fierz-Pauli mass terms are not present either.  
In such an unfamiliar setting a straightforward method 
to ascertain the degrees of freedom involves coupling to
sources~\cite{Salam:1981xd}.  
As shown in  
in appendix~\ref{App:DOF-two-der} the field $a_{ij}$ in the two-derivative approximation propagates:
\begin{enumerate}
\item  Ghost  spin-two  with  $m^2 = 4/\alpha'$.

\item  Ghost scalar with $m^2 = 4/\alpha'$.

\item  Scalar tachyon with $m^2 = -4/\alpha'$.

\end{enumerate}

\noindent 
The field $\bar a_{ij}$ in the two-derivative approximation propagates exactly
the same degrees
of freedom but with opposite value of mass-squared.

\subsection{Spectrum of the full six-derivative quadratic theory}

We now extend the above analysis to the full quadratic action including the higher derivative terms. 
Consider the four-derivative terms calculated before in (\ref{eq:L2426Final}).   
The signs in this expression are such that we can rewrite it as a difference of squares:
\be
\label{L42i}
 L^{(2,4)} \ =  \ 
 \tfrac{1}{16}\, (D_iD_j a^{ij}-{\cal R})^2
 -  \tfrac{1}{16}\,  (\bar D_i\bar D_j \bar a^{ij}-{\cal R})^2   \;. 
\ee 
Note now that the six-derivative terms in (\ref{eq:L2426Final}) are also of a similar form
\be \label{eq:L2426Final-vm}
  L^{(2,6)} \  = \
 \tfrac{1}{64}
 \bkt{ 
\bar{D}_i\bar{D}_j \bar{a}^{ij}\ 
+\ D_i D_j a^{ij} -2 {\cal R}}
\square \bkt{\bar{D}_i\bar{D}_j \bar{a}^{ij}\ 
+\ D_i D_j a^{ij} -2 {\cal R}}.
\ee
To see the structural form of the terms more clearly, define 
\be
x \equiv  \tfrac{1}{4} (D_iD_j a^{ij}-{\cal R})  \,, \quad  
\bar x  \equiv \tfrac{1}{4}  (\bar D_i\bar D_j a^{ij}-{\cal R})\,,
\ee
so that the full higher derivative Lagrangian can be written as
\be
\label{simple-rewrite}
L^{(2,4)}  + L^{(2,6)}  \ = \ x^2 - \bar x^2  \ + \tfrac{1}{4} (x + \bar x) \square ( x+ \bar x) \;. 
\ee
It is clear that the first two terms could be rewritten as a two-derivative Lagrangian with the
help of two auxiliary scalar fields $\varphi$ and $\bar \varphi$:
\be
\ x^2 - \bar x^2    \quad \to \quad    2 x \, \varphi  -  \varphi^2  \  - \ 2 \bar x\, \bar\varphi
 +  \bar\varphi^2  \ \equiv \  G (\varphi, \bar\varphi ) \;. 
\ee
One quickly sees that elimination of the auxiliary scalars leads to $\varphi = x$ and $\bar \varphi = \bar x$, 
giving back the terms to the left of the arrow. 
This means that 
\be
G (\varphi= x , \bar{\varphi}=\bar x )  \ = \ x^2 - \bar x^2 \,.
\ee
We also note the additional invariance property
\be
\label{adit}
G (\varphi = x + \eta  , \bar{\varphi}= \bar x - \eta  )  \ = \ x^2 - \bar x^2  \,. 
\ee
It is also clear that with the auxiliary field the Lagrangian
just has two derivatives.  
What is less obvious is that we 
can use the same idea for the full
higher-derivative Lagrangian in (\ref{simple-rewrite}).  We claim that   
\be
\label{simple-rewrite-vm}
L^{(2,4)}  + L^{(2,6)}  \ = \  
2 x \, \varphi  -  \varphi^2  \  - \ 2 \bar x\, \bar\varphi  
+  \bar\varphi^2 \ 
+ \tfrac{1}{4} (\varphi   + \bar \varphi  ) \square ( \varphi  + \bar \varphi  ) 
\ee
is on-shell fully equivalent to (\ref{simple-rewrite}). 
This is easily demonstrated.  The equations of motion for $\varphi$
and $\bar\varphi$ give
\be
\begin{split}
\varphi \ = \ & \ x + \tfrac{1}{4} \square (\varphi + \bar \varphi)\;, \\[0.5ex]
\bar \varphi \ = \ & \ x - \tfrac{1}{4} \square (\varphi + \bar \varphi)\;.  
\end{split}
\ee
It follows that $\varphi + \bar \varphi  = x+ \bar x$, which can be used for the last term
in (\ref{simple-rewrite-vm}).   Moreover, the above solution has the structure $\varphi= x + \eta$
and $\bar\varphi= \bar x - \eta$ so that the first two groups of terms in  (\ref{simple-rewrite-vm}), 
which equal $G (\varphi, \bar\varphi )$,  
still reproduce the first two terms in (\ref{simple-rewrite}) upon eliminating $\varphi$ and $\bar\varphi$. 

This demonstrates that  (\ref{simple-rewrite-vm}) provides a two derivative
Lagrangian that is equivalent to the original six-derivative one.  
A little more explicitly, the Lagrangian can be written as 
 \be\label{firstorderLxxx}
 \begin{split}
  L \ = \ &-\varphi^2 -\tfrac{1}{2}\,D_ia^{ij}D_j\varphi-\tfrac{1}{2}\,\varphi\,{\cal R}\\[0.5ex]
   &+\bar{\varphi}^2+\tfrac{1}{2}\,\bar{D}_i\bar{a}^{ij}\bar{D}_{j}\bar{\varphi} +\tfrac{1}{2}\,\bar{\varphi}\,{\cal R} \\[0.5ex]
& + \tfrac{1}{4}(\varphi + \bar{\varphi})\square (\varphi + \bar{\varphi})\;.  \end{split} 
 \ee 
Including the original two-derivative terms, we have the full quadratic action
 \be\label{simplequadL}
  \begin{split}
   L \ = \ &\,
   \tfrac{1}{4} e^{ij} \square  e_{ij}
+\tfrac{1}{4} \bkt{D_i e^{ij}}^2 
+\tfrac{1}{4}\bkt{\bar{D}_j e^{ij}}^2  +e^{ij}D_i \bar{D}_j\phi
-\phi\, \square \phi \\[1ex]  
   & - \tfrac{1}{8}\,  a^{ij} \, \square \, a_{ij}  -\tfrac{1}{4} \bkt{D_{i}a^{ij}}^2
    -\tfrac{1}{2}\,D_ia^{ij}D_j\varphi +\tfrac{1}{4}\varphi\square\varphi  \ + \tfrac{1}{2}\,  a^{ij} a_{ij} -\varphi^2\\[1ex]
    &- \tfrac{1}{8}\,  \bar{a}^{ij} \, \square \, \bar{a}_{ij}  -\tfrac{1}{4} \bkt{\bar{D}_{i}\bar{a}^{ij}}^2
    +\tfrac{1}{2}\,\bar{D}_i\bar{a}^{ij}\bar{D}_{j}\bar{\varphi}+\tfrac{1}{4}\bar{\varphi}\square\bar{\varphi} 
   \  - \tfrac{1}{2}\,  \bar{a}^{ij} \bar{a}_{ij}+\bar{\varphi}^2\\[1ex]
   &+\tfrac{1}{2}\bar{\varphi}\square \varphi  \ 
   -\tfrac{1}{2}\,\varphi\,{\cal R} + \tfrac{1}{2}\,\bar{\varphi}\,{\cal R}\;. 
  \end{split}
 \ee   
Now the terms in the second and third lines 
are improved compared to the two-derivative Lagrangian (\ref{bz-L22}). They have 
the derivative terms needed  
for a proper kinetic term and also 
mass terms for the new `dilatons' $\varphi$ and $\bar{\varphi}$. 

\medskip

The above action is not diagonal:  it has a  
$\bar{\varphi}\square \varphi$ term    
and $\varphi-\bar{\varphi}$ is coupled to the 
original DFT fields via ${\cal R}$. 
It turns out, however, that the action can be completely diagonalized by  an exact field redefinition of the dilaton. 
We let   
 \be
  \phi \; \rightarrow \; \phi' \ \equiv \ \phi-\tfrac{1}{2}(\varphi-\bar{\varphi})\;, 
 \ee
leaving all other fields unchanged.  Note that this redefinition is local and exactly invertible; 
hence there is no danger of inducing infinitely many terms.  
Denoting the standard quadratic  DFT Lagrangian in the first line of (\ref{simplequadL}) by $L_{\rm DFT}$ we compute 
\be
 L_{\rm DFT}[e_{ij},\phi] \ = \ L_{\rm DFT}[e_{ij},\phi']+\tfrac{1}{2}(\varphi-\bar{\varphi}){\cal R}(e,\phi')
 -\tfrac{1}{4}(\varphi-\bar{\varphi})\square (\varphi-\bar{\varphi})\;, 
\ee
which is an exact relation. The only other appearance of $\phi$ in (\ref{simplequadL}) is in the third line, 
inside ${\cal R}$, for which one computes with (\ref{eq:LinScalarR}) 
 \be
  {\cal R}(e,\phi) \ = \ {\cal R}(e,\phi') -\square (\varphi-\bar{\varphi})\;,  
 \ee
which is also exact. Using these two relations in  the full action (\ref{simplequadL})
one obtains (suppressing all arguments different from $\phi$) 
 \be 
 \begin{split}
  L[\phi] \ &= \  L[\phi']+\tfrac{1}{2}(\varphi-\bar{\varphi}){\cal R}(e,\phi')
 -\tfrac{1}{4}(\varphi-\bar{\varphi})\square (\varphi-\bar{\varphi})
  +\tfrac{1}{2}(\varphi-\bar{\varphi})\square (\varphi-\bar{\varphi})\\[1ex]
  \ &= \  L[\phi']+\tfrac{1}{2}(\varphi-\bar{\varphi}){\cal R}(e,\phi') 
  +\tfrac{1}{4}\varphi\square \varphi + \tfrac{1}{4}\bar{\varphi}\square \bar{\varphi}
  -\tfrac{1}{2}\bar{\varphi}\square \varphi\;. 
 \end{split} 
 \ee
This cancels the coupling between $\varphi-\bar{\varphi}$ and ${\cal R}$ as well as the term
$\bar{\varphi}\square \varphi$, while changing the coefficients of the diagonal terms  ${\varphi}\square \varphi$
and $\bar{\varphi}\square \bar{\varphi}$ from $\tfrac{1}{4}$ to $\tfrac{1}{2}$. 
Thus, dropping finally the $^{\prime}$ from $\phi$, the theory is fully equivalent to 
  \be\label{evensimplerquadL}
  \begin{split}
   L \ = \ &\,
   \tfrac{1}{4} e^{ij} \square  e_{ij}
+\tfrac{1}{4} \bkt{D_i e^{ij}}^2 
+\tfrac{1}{4}\bkt{\bar{D}_j e^{ij}}^2  +e^{ij}D_i \bar{D}_j\phi
-\phi\, \square \phi \\[1ex]  
   & - \tfrac{1}{8}\,  a^{ij} \, \square \, a_{ij}  -\tfrac{1}{4} \bkt{D_{i}a^{ij}}^2
    -\tfrac{1}{2}\,D_ia^{ij}D_j\varphi +\tfrac{1}{2}\varphi\square\varphi  \ + \tfrac{1}{2}\,  a^{ij} a_{ij} -\varphi^2\\[1ex]
    &- \tfrac{1}{8}\,  \bar{a}^{ij} \, \square \, \bar{a}_{ij}  -\tfrac{1}{4} \bkt{\bar{D}_{i}\bar{a}^{ij}}^2
    +\tfrac{1}{2}\,\bar{D}_i\bar{a}^{ij}\bar{D}_{j}\bar{\varphi}+\tfrac{1}{2}\bar{\varphi}\square\bar{\varphi} 
   \  - \tfrac{1}{2}\,  \bar{a}^{ij} \bar{a}_{ij}+\bar{\varphi}^2 \;, 
  \end{split}
 \ee
which is now diagonal, so that we can readily study the physical content. 
The analysis in appendix~\ref{a2app} shows 
that the fields $(a_{ij}, \varphi)$  propagate:
 
\noindent 
\begin{enumerate}

\item  Ghost  spin-two  with  $m^2 = 4/\alpha'$.

\item  Ghost scalar with $m^2 = 4/\alpha'$.

\item  Scalar with $m^2 = 4/\alpha'$.

\end{enumerate}  

\noindent 
These are the same degrees of freedom as in the two-derivative approximation, except
that the scalar tachyon turned into a healthy massive scalar. 
The fields $(\bar a_{ij}, \bar\varphi)$  propagate exactly
the same degrees
of freedom as the un-barred pair but with opposite value of mass-squared.

\medskip
We conclude by noting that a further redefinition
of $\varphi$ and the trace $a$ of $a_{ij}$  allows us 
to fully diagonalize into massive spin-2 and a massive scalar in the
Lagrangian (\ref{evensimplerquadL}). 
We let     
 \be
  \varphi \ = \ \varphi'-\tfrac{1}{2}a'\;, \qquad
  a_{ij} \ = \ a_{ij}'-\varphi' \eta_{ij}\;. 
 \ee 
Inserting this into the second line of the action above and dropping  primes at 
the end, one obtains 
 \be\label{diagMAss}
 \begin{split}
  L \ = \  &\; - \tfrac{1}{8} a^{ij} \, \square \, a_{ij}  -\tfrac{1}{4} (D_{i}a^{ij})^2
  -\tfrac{1}{4}a^{ij}\,D_iD_{j}a    
  +\tfrac{1}{8}a\square a 
  +\tfrac{1}{2}\big(a^{ij} a_{ij}-\tfrac{1}{2}a^2\big)\\[1ex]
  &-\tfrac{1}{4}(D-2)\big(\tfrac{1}{2}\varphi\square\varphi-2\varphi^2\big)\;. 
 \end{split} 
 \ee  
The second line implies that $\varphi$ is a ghost with mass $M^2=4$. 
The first line has the right kinetic terms as in the Fierz-Pauli theory, but the mass term has the wrong 
relative 
coefficient.\footnote{Curiously, the mass term obtained here coincides with the `mass term' 
obtained by expanding a cosmological constant term proportional to $\sqrt{|g|}$ around flat space, 
c.f.~\cite{tHooft:2007bf}} 
Thus, in addition to the (ghostly) massive spin-2 it propagates 
a scalar mode, given by the trace $a$.

\sectiono{Massive linearized DFT}  \label{masslindft}

 The linearized DFT action describes massless gravity,
a massless two-form field, and a massless dilaton.  We find here a duality invariant
mass term that gives the same mass to all these three fields, without introducing
ghosts or spurious degrees of freedom.   
For linearized Einstein gravity a consistent massive deformation requires a 
judicious choice of mass terms:  the Fierz-Pauli mass term, 
that involves both the trace $h^{ij} h_{ij}$ of the square of the metric fluctuation
and the square of the trace $h$.  The latter is required to 
guarantee that $h$ is non-propagating, for otherwise it would be 
a scalar ghost. 
In DFT, the trace of the field $e_{ij}$ is not available because there is no $O(D,D)$
covariant notion of taking this trace, 
but one can give a novel mass term involving the dilaton, which also avoids all scalar ghosts.

Consider the linearized two-derivative DFT action, given on the first line in 
(\ref{bz-L22}):
\be
\label{bz-L22-new}  
\begin{split}
 L_{\text{DFT}}  \  = \ \, & \tfrac{1}{4} e^{ij} \square e_{ij}
+\tfrac{1}{4} \bkt{D_i e^{ij}}^2 
+\tfrac{1}{4}\bkt{\bar{D}_j e^{ij}}^2 
+e^{ij}D_i \bar{D}_j\phi \ -\phi\, \square \phi  \\[0.5ex]
\ = \ \,&  \tfrac{1}{2}e^{ij}{\cal R}_{ij}(e,\phi)+\tfrac{1}{2}\phi\, {\cal R}(e,\phi) \,,
\end{split}
\ee
where we rewrote the kinetic terms geometrically, 
discarding total derivatives, 
 in terms of the linearized Ricci tensor ${\cal R}_{ij}$ and the scalar curvature ${\cal R}$ 
 defined in (\ref{eq:LinScalarR}).\footnote{Note that the total variation 
 takes the form 
 $\delta L_{\rm DFT} \  =  \ \delta e^{ij}\, {\cal R}_{ij} +\delta \phi \,{\cal R}$, 
 discarding total derivatives as usual.} 
We add to this linearized two-derivative DFT action mass terms in the following way:  
\be\label{massiveDFT}  
  L_{\text{mDFT}} \ = \  
    \tfrac{1}{2}e^{ij}{\cal R}_{ij}(e,\phi)+\tfrac{1}{2}\phi{\cal R}(e,\phi) 
            -\tfrac{1}{4}M^2(e^{ij} e_{ij}-  4 \phi^2)\;.
 \ee         
Note that $O(D,D)$ covariance does not restrict the relative coefficient between the 
mass terms of $e_{ij}$ and the dilaton $\phi$, but we will show in the following 
that the specific choice made here leads to a ghost-free model. 
One way to see this is to inspect the field equations for $e_{ij}$ and $\phi$,   
 \be\label{EOMmass0}
 \begin{split}
   {\cal R}_{ij} \ &= \ \tfrac{1}{2}M^2 e_{ij}\, ,\qquad 
   {\cal R} \ = \ -2 M^2 \phi\;. 
 \end{split}
 \ee     
The generalized Ricci tensor and scalar curvature
satisfy the Bianchi identities 
 \be
  D^i {\cal R}_{ij} \ = \ -\tfrac{1}{2}\bar{D}_j{\cal R}\;, \qquad \bar{D}^j{\cal R}_{ij} \ =  \ -\tfrac{1}{2}D_i{\cal R}\;, 
 \ee
so that taking the divergence and derivative of the field equations we obtain 
  \be\label{divSTEP}
  \begin{split}
   0 \ = \ D^i {\cal R}_{ij}+\tfrac{1}{2}\bar{D}_j{\cal R} \ = \    \tfrac{1}{2}M^2 (D^ie_{ij}-2 \bar{D}_j\phi)\,, \\[0.5ex]
 0 \ = \ \bar D^j {\cal R}_{ij}+\tfrac{1}{2}\bar{D}_i{\cal R} \ = \    \tfrac{1}{2}M^2 (\bar D^je_{ij}-2{D}_i\phi)\,.   
   \end{split}
  \ee 
Taking another divergence, this implies
 \be
  D^i\bar{D}^j e_{ij}-2\square \phi \ = \ 0 \quad \Rightarrow \quad {\cal R} \ = \ 0\;, 
 \ee
where we used the explicit expression for the scalar curvature.  
Thus, thanks to the specific choice of mass terms, the scalar curvature vanishes on-shell, 
which in turn removes propagating degrees of freedom that would otherwise be present. 
Indeed, from this we conclude with  (\ref{EOMmass0}) that $\phi=0$ and hence with (\ref{divSTEP}) that both barred and unbarred divergences of $e_{ij}$ vanish on-shell: 
 \be\label{massDFTconstr}
  D^i e_{ij} \ = \ \bar{D}^j e_{ij} \ = \ \phi \ = \ 0\;. 
 \ee  
This should be compared to on-shell constraints of the Fierz-Pauli theory for massive (linearized) gravity, which are 
$\partial^{\mu} h_{\mu\nu}=0$ and $h^{\mu}{}_{\mu}=0$, and the on-shell constraint of the massive two-form field, 
which is $\partial^{\mu}b_{\mu\nu}=0$.  
We note that (\ref{massDFTconstr}) gives as many 
constraints as needed in order to describe a massive graviton, a massive two-form field, and a massive 
scalar. Indeed, with the on-shell constraints the field equation becomes
$( \square - M^2 ) e_{ij}=0$ and, in a frame where $p_\mu= (M, \vec{0})$, we see
that $e_{0i} = e_{i0} =0$, resulting in $(D-1)^2$ degrees of freedom describing
a graviton, a two-form field and a scalar, all of mass $M$.
Interestingly, in DFT variables the massive scalar is \textit{not} encoded in the dilaton density $\phi$, 
which vanishes on-shell, but rather in the trace of $e_{ij}$, which can only be accessed after breaking manifest 
$O(D,D)$ covariance. 
It should also be noted that although the kinetic terms of massive DFT can be diagonalized 
(after abandoning manifest $O(D,D)$ invariance), this field redefinition does not diagonalize the mass terms. 
Therefore, this model is not simply the 
Fierz-Pauli theory of massive gravity 
supplemented by  
a massive 2-form and a massive scalar.

It is instructive to make this point a little more explicit. Since the $b$-field plays no role in this 
discussion, we will set it to zero and, having thus abandoned $O(D,D)$ invariance, denote
the spacetime indices by $\mu,\nu,\ldots$,
take the derivatives $D$ and $\bar D$ to be
partial derivatives and $\square = \partial^2$.  The Lagrangian (\ref{massiveDFT})  
then gives:
\be \label{eq:mDFTnoDuality}
L\ =\  \tfrac{1}{4} h^{\mu\nu} \square h_{\mu\nu}
+\tfrac{1}{2} \bkt{\p_{\mu} h^{\mu\nu}}^2 
+h^{\mu\nu}\p_{\mu} \p_{\nu}\phi \ -\phi\, \square \phi\ -\tfrac{1}{4} M^2\bkt{h^{\mu\nu} h_{\mu\nu}\ - 4\phi^2}\,.   
\ee
We want to see if this is field redefinition equivalent to the Fierz-Pauli action supplemented
by a massive scalar:
\be
\begin{split}
L_{FP} + L_s\ =\ & \  \tfrac{1}{4} h^{\mu\nu} \square h_{\mu\nu}
+\tfrac{1}{2} \bkt{\p_{\mu} h^{\mu\nu}}^2 
+\tfrac{1}{2} \, h^{\mu\nu}\p_{\mu} \p_{\nu}h \ -\tfrac{1}{4} h\, \square h\ -\tfrac{1}{4} M^2\bkt{h^{\mu\nu} h_{\mu\nu}\ - h^2}    \\[1.0ex]
& \ +  \phi\, \square \phi - M^2 \phi^2 \,. 
\end{split}
\ee 
The most general field redefinition can be parameterized as follows:
\be \label{eq:mDFTRed}
h_{\mu\nu}\ = \ A_{1} h'_{\mu\nu}\ + \ \eta_{\mu\nu} \bkt{A_2 h' \ + A_3 \phi '}\; , \qquad \phi\ =\  A_4 h'\ + A_5 \phi'.
\ee
For this field redefinition to be invertible
 $A_1$ has to be non-zero. We will now show that there is no choice of coefficients $A_1,\, \cdots ,A_5$ that define an invertible redefinition
and {\em simultaneously} diagonalize the 
kinetic and mass terms of massive DFT. Using  (\ref{eq:mDFTRed}) in the  Lagrangian (\ref{eq:mDFTnoDuality}) we get:
\be 
\begin{split}
L = &
\  A_1 \bkt{- A_3 +  A_5 } {h'}^{\mu\nu}\p_{\mu}\p_{\nu} \phi'\ 
+ \bkt{  \tfrac{1}{2} A_3\bkt{A_1\ + \bkt{D-2}A_2 \ + 2A_4 }+ A_2 A_5 -2A_4A_5} h'\square \phi'\ \\[1ex]
&-\tfrac{1}{2} M^2\,  \bkt{ A_1A_3 \ + A_2A_3 D- 4A_4 A_5 } h'\phi'\ + \cdots ,\label{eq:LmDFTRed}
\end{split}
\ee
where `$\cdots$' indicates diagonal terms. Requiring the 
off-diagonal terms to vanish, we find two solutions 
\be 
A_3 \ = \ 0 \ = \ A_5\; ,  \qquad  \text{or }\qquad A_2\ =\ -\frac{A_1}{D}\; , \qquad A_3\ = \ A_5 , \qquad  A_4 \ = \ 0\;.
\ee
In the first solution the field redefinition (\ref{eq:mDFTRed}) does not involve $\phi'$ and hence is not invertible.  With the second solution,  
 \be 
 h_{\mu\nu}\ = \ A_1\bkt{{h}'_{\mu\nu}\ -\ \tfrac{1}{D} \eta_{\mu\nu}h'}\ + \  A_3\eta_{\mu\nu} \phi'\; , \qquad \phi\ =\ A_3\phi' .
 \ee
Only the traceless part of $h'_{\mu\nu}$ appears and hence the redefinition is non-invertible. We conclude that there is no field redefinition which diagonalizes {\em both}
 the kinetic and the mass term of  massive DFT.

Let us now perform a field redefinition which diagonalizes the kinetic term of massive DFT:
\be 
  h_{\mu\nu} \ = \ h_{\mu\nu}'+\phi' \eta_{\mu\nu}\;,  \qquad 
  \phi \ = \ \phi'+\tfrac{1}{2}h'\;,
\ee
after which the Lagrangian (\ref{eq:mDFTnoDuality}),  
upon dropping primes, reads
 \be\label{diagLag}
  \begin{split}
   L \ = \ &\, -\tfrac{1}{2}h^{\mu\nu} G_{\mu\nu}(h) \ - \ \tfrac{1}{4}M^2(h^{\mu\nu}h_{\mu\nu}-h^2)\\[1ex]
   &\, +\tfrac{1}{4}(D-2)\phi\square \phi \ 
   - \ \tfrac{1}{4}(D-4)M^2\phi^2 \ - \ \tfrac{1}{2}M^2\phi h\;.   
  \end{split}
 \ee   
Here $G_{\mu\nu}(h)$ is the linearized Einstein tensor,   
\be
\label{lin-Einstein-ten}
G_{\mu\nu}(h) \ =\ R_{\mu\nu}(h) -\tfrac{1}{2}R(h) \eta_{\mu\nu}\, , 
\ee
where the linearized Ricci tensor and scalar curvatures are  
 \be\label{linRicci}
  R_{\mu\nu}(h) \ = \ -\tfrac{1}{2}(\square h_{\mu\nu}
  -2\partial_{(\mu}\partial^{\rho} h_{\nu)\rho}+\partial_{\mu}\partial_{\nu}h) \,, \quad 
  R(h) \ = \ - \square h + \partial^\mu\partial^\nu h_{\mu\nu} \,. 
 \ee
 Under the integral one quickly checks that  
  \be
 -\tfrac{1}{2}h^{\mu\nu} G_{\mu\nu}(h) \ = \ \tfrac{1}{4} h^{\mu\nu} \square h_{\mu\nu}
+\tfrac{1}{2} \bkt{\p_{\mu} h^{\mu\nu}}^2 
+\tfrac{1}{2} \, h^{\mu\nu}\p_{\mu} \p_{\nu}h \ -\tfrac{1}{4} h\, \square h\,. 
 \ee
The linearized Einstein tensor is self-adjoint: under an integral, $a^{\mu\nu} G_{\mu\nu}(b)=b^{\mu\nu} G_{\mu\nu}(a)$, for arbitrary symmetric 
tensors $a$ and $b$.

As claimed,   
the kinetic terms in (\ref{diagLag}) 
are now diagonal, but the mass terms contain
the non-removable 
term $\phi h$. Nevertheless, it is straightforward to see that this model propagates the 
right number of degrees of freedom by analyzing the field equations, which read 
 \be\label{massiveFieldEQ}
  \begin{split}
   G_{\mu\nu}(h) \ &= 
   \ -\tfrac{1}{2}M^2(h_{\mu\nu}-h\eta_{\mu\nu}+\phi \eta_{\mu\nu})\,,   \\[1ex]
   \tfrac{1}{2}(D-2)\square \phi \ &= \ \tfrac{1}{2}(D-4)M^2\phi +\tfrac{1}{2}M^2 h\;. 
  \end{split}
 \ee  
Taking the divergence of the first equation and using the Bianchi identity $\partial^{\mu} G_{\mu\nu}=0$ 
implies 
 \be\label{constraintsubsid} 
  \partial^{\mu} h_{\mu\nu} - \partial_{\nu} h \ = \ -\partial_{\nu}\phi\;, 
 \ee
which after taking another divergence and using the explicit expression for the Ricci scalar $R(h)$ implies 
 \be\label{Requalsbox}
  R(h) \ = \ -\square \phi\;. 
 \ee
Taking the trace of the first equation in (\ref{massiveFieldEQ}) we thus obtain 
 \be
  -\tfrac{1}{2}(D-2)R(h) \ = \ \tfrac{1}{2}(D-2)\square \phi \ = \ \tfrac{1}{2}M^2(D-1) h -\tfrac{1}{2}M^2 D \phi\;. 
 \ee  
Together with the second equation in (\ref{massiveFieldEQ}) this implies  
 \be
  0 \ = \ \tfrac{1}{2}M^2(D-2)(h-2\phi) \qquad \Rightarrow \qquad h \ = \ 2\phi\;, 
 \ee 
assuming $D>2$.  
 Using $h=2\phi$ in the second equation in (\ref{massiveFieldEQ}) finally yields 
  \be\label{KGphi}
   (\square - M^2)\phi \ = \ 0\;, 
  \ee
proving that $\phi$ propagates non-tachyonically with mass $M$.  Note that this physical mass 
differs from the naive mass read off from the non-diagonal Lagrangian (\ref{diagLag}).  
Using now (\ref{KGphi}), (\ref{Requalsbox}) and $h=2\phi$ in the first equation in (\ref{massiveFieldEQ}) 
one obtains $R_{\mu\nu}(h) \ = \ -\tfrac{1}{2}M^2 h_{\mu\nu}$. Using the constraint (\ref{constraintsubsid}) 
together with $h=2\phi$ in the explicit expression for $R_{\mu\nu}(h)$, one finally obtains 
 \be
  (\square - M^2)h_{\mu\nu} \ = \ 0\;, 
 \ee
proving that $h_{\mu\nu}$ propagates as a massive spin-2 mode of mass $M$.  
Since both $h_{\mu\nu}$ and $\phi$ have the right-sign kinetic terms in (\ref{diagLag}) 
the model propagates precisely the expected (healthy) massive modes.

As a further check appendix \ref{App:mDFT} gives a source analysis
of the model. 
The results confirm
 that massive DFT describes a massive graviton, massive dilaton, and massive 2-form field and does not propagate any undesired (ghost-like) modes.  
 We also
 show that the particular combination $e^{ij} e_{ij} - 4 \phi^2$ of mass
 terms is strictly necessary:  for any other combination one finds an additional
 ghost scalar.

 \sectiono{Tensionless limit 
  and degrees of freedom}  \label{tenlimdegfre}

We  revisit the 
HSZ quadratic theory 
in a form for which one can take the tensionless
limit and see the appearance of an enhanced gauge symmetry.
The addition of  higher-derivative terms generally increases 
the number of propagating degrees of freedom, 
typically leading to ghost modes,
but in HSZ   theory the number of modes is unchanged. 
We give here an alternative model  where the addition of higher-derivative terms 
to a two-derivative action \textit{reduces} the number of degrees of freedom. 
Such reduction  hinges on a set of Bianchi identities that govern the derivative structure
of the theory and that are a consequence of the enhanced gauge invariance in the tensionless limit $\alpha'\rightarrow \infty$.

\subsection{Tensionless limit, enhanced gauge symmetry, and Bianchi identities }

Recall the full quadratic Lagrangian (\ref{evensimplerquadL}) for HSZ theory:
  \be\label{evensimplerquadL0}
  \begin{split}
   L \ = \ &\,
   \tfrac{1}{4} e^{ij} \square  e_{ij}
+\tfrac{1}{4} \bkt{D_i e^{ij}}^2 
+\tfrac{1}{4}\bkt{\bar{D}_j e^{ij}}^2  +e^{ij}D_i \bar{D}_j\phi
-\phi\, \square \phi \\[1ex]  
   & - \tfrac{1}{8}\,  a^{ij} \, \square \, a_{ij}  -\tfrac{1}{4} \bkt{D_{i}a^{ij}}^2
    -\tfrac{1}{2}\,D_ia^{ij}D_j\varphi +\tfrac{1}{2}\varphi\square\varphi  \ + \ \tfrac{1}{\alpha'}(\tfrac{1}{2}\,  a^{ij} a_{ij} -\varphi^2)\\[1ex]
    &- \tfrac{1}{8}\,  \bar{a}^{ij} \, \square \, \bar{a}_{ij}  -\tfrac{1}{4} \bkt{\bar{D}_{i}\bar{a}^{ij}}^2
    +\tfrac{1}{2}\,\bar{D}_i\bar{a}^{ij}\bar{D}_{j}\bar{\varphi}+\tfrac{1}{2}\bar{\varphi}\square\bar{\varphi} 
   \  - \ \tfrac{1}{\alpha'}( \tfrac{1}{2}\,  \bar{a}^{ij} \bar{a}_{ij}-\bar{\varphi}^2 ) \;, 
  \end{split}
 \ee   
where we restored the $\alpha'$ dependence, which is fixed by dimensional analysis. 
In this form, the limit $\alpha'\rightarrow \infty$ can be taken smoothly, which simply sets the mass terms to zero. 
For this limiting Lagrangian we define the variations w.r.t.~$a$, $\bar{a}$, 
$\varphi$, and $\bar{\varphi}$, respectively, 
 \be\label{DEFvar}
  \delta L_{\alpha'\rightarrow \infty} \ = \ \delta a^{ij} {\cal A}_{ij}+\delta \bar{a}^{ij}\bar{\cal A}_{ij}
  +\delta\varphi\, {\cal S} + \delta\bar{\varphi}\,\bar{\cal S}\;. 
 \ee 
The variations, viewed as occurring under an integral, give
 \be\label{ASDEF}
  \begin{split}
   {\cal A}_{ij} \ &= \ -\tfrac{1}{4}\square a_{ij}+\tfrac{1}{2}D_{(i}D^k a_{j)k}+\tfrac{1}{2}D_iD_j\varphi\;, \qquad
    {\cal S} \ = \ \square \varphi+\tfrac{1}{2}D^i D^ja_{ij} \;, \\[1ex]
   \bar{\cal A}_{ij} \ &= \ -\tfrac{1}{4}\square \bar{a}_{ij}+\tfrac{1}{2}\bar{D}_{(i}\bar{D}^k \bar{a}_{j)k}
   -\tfrac{1}{2}\bar{D}_i\bar{D}_j\bar{\varphi}\;, \qquad 
    \bar{\cal S} \ = \ \square \bar{\varphi} - \tfrac{1}{2}\bar{D}^i \bar{D}^j\bar{a}_{ij}\;. 
  \end{split}
 \ee  
These tensors are the analogues of the Ricci tensor and curvature scalar of the usual DFT, 
with the difference that ${\cal A}_{ij}$ carries unbarred/unbarred indices 
and $\bar{\cal A}_{ij}$  carries barred/barred indices.  
It is easy to verify that they satisfy the Bianchi identities 
 \be\label{ASBianchis}
  \begin{split}
   & D^i{\cal A}_{ij}-\tfrac{1}{2}D_j{\cal S} \ = \ 0\;, \\[1ex]
   & \bar{D}^i\bar{\cal A}_{ij} + \tfrac{1}{2}\bar{D}_{j}\bar{\cal S} \ = \ 0 \;.
  \end{split}
 \ee  
It then immediately follows with (\ref{DEFvar}) that the Lagrangian in the `tensionless limit' 
$\alpha'\rightarrow \infty$ exhibits the enhanced gauge symmetry 
 \be\label{gaugeNEW}
 \begin{split}
  \delta_{\zeta}a_{ij} \ &= \ D_{i}\zeta_j+D_j\zeta_i\;, \qquad
  \delta_{\zeta}\varphi \ = \  -D_i\zeta^i\;, \\[1ex]
  \delta_{\bar{\zeta}}\bar{a}_{ij} \ &= \ \bar{D}_{i}\bar{\zeta}_j+\bar{D}_j\bar{\zeta}_i\;, \qquad
  \delta_{\bar{\zeta}}\bar{\varphi} \ = \  \bar{D}_i\bar{\zeta}^i \;, 
 \end{split}
 \ee 
with independent gauge parameters $\zeta^i$ and $\bar{\zeta}^i$, 
where the massless DFT fields $e_{ij}$ 
and $\phi$ stay invariant.\footnote{Note, however, that since the dilaton entering here is really 
the redefinition $\phi'=\phi-\tfrac{1}{2}(\varphi-\bar{\varphi})$ of the original dilaton $\phi$, it follows  
that the latter transforms as $\delta\phi=-\tfrac{1}{2}(D_i\zeta^i+\bar{D}_{i}\bar{\zeta}^i)$ under the new symmetry.}
It is amusing to compare this with the gauge symmetry of the original DFT, which  
has gauge parameters $\lambda^i$ and $\bar{\lambda}^i$ of the same form. These encode (linear combinations of) linearized diffeomorphisms and the 
 gauge symmetry of the two-form field, but the gauge invariance discovered here does not seem to have 
 an interpretation in terms of conventional (spacetime) symmetries.

\subsection{Higher derivatives that reduce the number of degrees of freedom}

We now demonstrate that adding higher-derivative terms 
to an action can \textit{reduce} the number of propagating degrees of freedom. As we have seen in section 3, 
the quadratic approximation to HSZ theory is already quite special in that the addition of 
higher-derivative terms does \textit{not} increase the number of degrees of freedom. 
The reason for this can be traced to the rewriting of higher-derivative 
contributions as two-derivative terms by means of extra fields, leading to an improved structure of the kinetic terms, as made clear by the emergence of a gauge invariance in the massless limit.   
With this improvement there is actually one choice for the mass terms that would 
have reduced the number of degrees of freedom. 
Although this is \textit{not} HSZ theory, it is interesting in its own right.
To our knowledge, this  
phenomenon was not known before in the literature. 

Specifically, let us consider adding to the same two-derivative Lagrangian 
(\ref{bz-L22})  the alternative four- and six-derivative terms 
 \be\label{alternativeL}
 \begin{split}
  L^{(4,2)}_{\text{alternative}} \ &= \ \tfrac{1}{32}(D_iD_ja^{ij}-{\cal R})^2 - 
  \tfrac{1}{32}(\bar{D}_i\bar{D}_j\bar{a}^{ij}-{\cal R})^2\;, \\[1ex]
  L^{(6,2)}_{\text{alternative}}  \ &= \ \tfrac{1}{4\cdot 64}(D_iD_ja^{ij}+\bar{D}_i\bar{D}_j\bar{a}^{ij}-2{\cal R})\square 
  (D_iD_ja^{ij}+\bar{D}_i\bar{D}_j\bar{a}^{ij}-2{\cal R})\;. 
 \end{split}
 \ee
 Comparing with (\ref{L42i}) and (\ref{eq:L2426Final-vm}) 
we see that these have precisely the same structure as in 
HSZ theory,  but with relative coefficients of $\frac{1}{2}$ 
for the four-derivative terms and $\frac{1}{4}$ for the six-derivative terms. 
As before, we can now pass to a formulation that is second-order in derivatives by introducing 
auxiliary fields $\varphi$ and $\bar{\varphi}$, so that we obtain for the full Lagrangian 
 \be
 \begin{split}
  L_{\text{alternative}} \ = \ & \ \  L^{(2, \leq 2)} 
   -2\varphi^2 +\tfrac{1}{2} \varphi(D_iD_ja^{ij} -{\cal R}) 
   +2\bar{\varphi}^2-\tfrac{1}{2}\bar\varphi(\bar{D}_i\bar{D}_j\bar{a}^{ij}-{\cal R})
   \\[0.5ex]
   &+\tfrac{1}{4}(\varphi + \bar{\varphi})\square (\varphi + \bar{\varphi})\;, 
    \end{split} 
 \ee 
where $L^{(2, \leq 2)}$ denotes the complete quadratic DFT Lagrangian in (\ref{bz-L22}).  
Integrating out $\varphi$ and $\bar{\varphi}$ one recovers the higher-derivative terms in (\ref{alternativeL}).   This alternative Lagrangian is the same Lagrangian as in 
(\ref{simplequadL}) except for the mass terms for $\varphi$ and $\bar \varphi$.
Next, we perform the same field redefinition 
$\phi  \rightarrow  \phi'  \equiv  \phi-\tfrac{1}{2}(\varphi-\bar{\varphi})$ of the dilaton used earlier
to obtain (\ref{evensimplerquadL}).  This time we get: 
   \be
  \begin{split}
   L_{\rm alternative} \ = \ &\, \ \ 
   \tfrac{1}{4} e^{ij} \square  e_{ij}
+\tfrac{1}{4} \bkt{D_i e^{ij}}^2 
+\tfrac{1}{4}\bkt{\bar{D}_j e^{ij}}^2  +e^{ij}D_i \bar{D}_j\phi
-\phi \square \phi \\[1ex]  
   & - \tfrac{1}{8}\,  a^{ij} \, \square \, a_{ij}  -\tfrac{1}{4} \bkt{D_{i}a^{ij}}^2
    -\tfrac{1}{2}\,D_ia^{ij}D_j\varphi +\tfrac{1}{2}\varphi\square\varphi  \ +  \ \tfrac{1}{\alpha'}
    (\tfrac{1}{2}\,  a^{ij} a_{ij} -2\varphi^2)\\[1ex]
    &- \tfrac{1}{8}\,  \bar{a}^{ij} \, \square \, \bar{a}_{ij}  -\tfrac{1}{4} \bkt{\bar{D}_{i}\bar{a}^{ij}}^2
    +\tfrac{1}{2}\,\bar{D}_i\bar{a}^{ij}\bar{D}_{j}\bar{\varphi}+\tfrac{1}{2}\bar{\varphi}\square\bar{\varphi} 
   \  - \ \tfrac{1}{\alpha'}(\tfrac{1}{2}\,  \bar{a}^{ij} \bar{a}_{ij}-2\bar{\varphi}^2) \;. 
  \end{split}
 \ee   
This only differs from (\ref{evensimplerquadL}) 
in the coefficients of the mass terms for $\varphi$ and $\bar{\varphi}$. 

We will now show that, thanks to the precise coefficients of the mass terms, the number of 
degrees of freedom is \textit{reduced} compared to the original two-derivative theory. 
This analysis is largely analogous to that of the massive DFT model in sec.~\ref{masslindft}. 
We first consider the field equations for $a$, $\varphi$ and  $\bar{a}$, $\bar{\varphi}$,   
 \be\label{aphiEQ0}  
 \begin{split}
  {\cal A}_{ij} +\tfrac{1}{\alpha'}a_{ij} \ = \ 0\;, & \qquad
  {\cal S}-\tfrac{4}{\alpha'}\varphi \ = \ 0\;, \\[0.5ex]
   \bar {\cal A}_{ij} -\tfrac{1}{\alpha'}\bar a_{ij} \ = \ 0\;, & \qquad
 \bar {\cal S}+\tfrac{4}{\alpha'}\bar\varphi \ = \ 0\;, \\
  \end{split}
 \ee
where the tensors are defined in (\ref{ASDEF}).  
Taking divergence and derivative of these equations, we infer with the  Bianchi identities in (\ref{ASBianchis}) 
  \be\label{SINgDIV}
  \begin{split}
   0 \ = \ D^i{\cal A}_{ij}-\tfrac{1}{2}D_j{\cal S} \quad \Rightarrow \quad
   D^i a_{ij}+2 D_j\varphi \ = \ 0\, , \\[0.5ex]
   0 \ = \ \bar D^i \bar {\cal A}_{ij}+\tfrac{1}{2}\bar D_j\bar {\cal S} \quad \Rightarrow \quad
   \bar D^i \bar a_{ij}-2 \bar D_j\bar \varphi \ = \ 0\, .
   \end{split}
  \ee
Taking another divergence we obtain 
 \be
 \begin{split}
   D^iD^j a_{ij}+2 \square \varphi \ = \ 0   & \ 
   \quad \Rightarrow \quad {\cal S} \ = \ 0\;,  \\[0.5ex]
    \bar D^i \bar D^j \bar a_{ij}-2 \square \bar\varphi \ = \ 0   & \ 
   \quad \Rightarrow \quad \bar {\cal S} \ = \ 0\;,  
 \end{split}
 \ee
where we used the explicit forms of ${\cal S}$ and $\bar{\cal S}$ in (\ref{ASDEF}). 
Back in (\ref{aphiEQ0}) this implies $\varphi= \bar \varphi= 0$ and so with (\ref{SINgDIV})  
 \be
  D^i a_{ij} \ = \ \bar{D}^i \bar{a}_{ij} \ = \ 0\;, \qquad \varphi \ = \ \bar\varphi \ = \ 0\;. 
 \ee
Thus, we obtained as many constraints as needed in order to describe precisely
two massive spin-2 fields and two massive scalars. Indeed, using these constraints 
in the equations of motion (see (\ref{aphiEQ0})), 
the dynamical equations become 
 \be
  \big(\square -\tfrac{4}{\alpha'}\big)a_{ij} \ = \ 0\;, \qquad 
  \big(\square +\tfrac{4}{\alpha'}\big)\bar{a}_{ij} \ = \ 0\;, 
 \ee
which propagate the spin-2 and spin-0 parts of $a_{ij}$ with mass $m^2=\frac{4}{\alpha'}$, 
while the spin-2 and spin-0 parts of $\bar{a}_{ij}$ are propagated tachyonically but with the same mass-squared. 
Moreover, the source analysis for massive DFT applies immediately to the present model, the only difference being that here we have two massive spin-2 fields and no massive two-form fields. Summarizing, the total content of massive 
states is two massive spin-2 and two massive scalars. Hence, compared to the original two-derivative 
model, we lost two scalar modes upon adding higher-derivative terms.

\sectiono{Eliminating the massive fields}

We have seen that the quadratic theory propagates the familiar massless
degrees of freedom as well as massive and ghost-like spin-2 and spin-0 
degrees of freedom. In 
 this section we will show that by a simple 
redefinition the massive fields 
can be rendered truly auxiliary.  The redefinition sacrifices manifest T-duality,
but allows us to eliminate these fields  
algebraically, leading to an infinite series of higher-derivative corrections for the massless fields. 
To any finite order in $\alpha'$ these higher-derivative terms can be removed by a local field redefinition,  but removing them to all orders in $\alpha'$  would require an illegal non-local redefinition.

Our goal is to eliminate the massive fields from the quadratic Lagrangian (\ref{evensimplerquadL}).  
The method used in this  
requires giving up manifest $O(D,D)$ covariance,  and accordingly we write spacetime 
indices as $\mu,\nu, \ldots$ and take the derivatives $D$ and $\bar D$ to be
partial derivatives and $\square = \partial^2$:
\be\label{evensimplerquadLvmbb1}
  \begin{split}
   L \ = \ &\,
   \tfrac{1}{4} e^{\mu\nu} \square  e_{\mu\nu}
+\tfrac{1}{4} \bkt{\partial_\mu e^{\mu\nu}}^2 
+\tfrac{1}{4}\bkt{\partial_\nu e^{\mu\nu}}^2  +e^{\mu\nu}\partial_\mu 
\partial_\nu\phi
-\phi\, \square \phi \\[1ex]  
   & - \tfrac{1}{8}\,  a^{\mu\nu} \, \square \, a_{\mu\nu}  -\tfrac{1}{4} 
   \bkt{\partial_\mu a^{\mu\nu}}^2
    -\tfrac{1}{2}\,\partial_\mu a^{\mu\nu}\partial_\nu\varphi 
    +\tfrac{1}{2}\varphi\square\varphi  \ + \tfrac{1}{2}\,  a^{\mu\nu} a_{\mu\nu} -\varphi^2\\[1ex]
    &- \tfrac{1}{8}\,  \bar{a}^{\mu\nu} \, \square \, \bar{a}_{\mu\nu}  -\tfrac{1}{4} \bkt{\partial_\mu\bar{a}^{\mu\nu}}^2
    +\tfrac{1}{2}\,\partial_\mu\bar{a}^{\mu\nu}\partial_\nu
    \bar{\varphi}+\tfrac{1}{2}\bar{\varphi}\square\bar{\varphi} 
   \  - \tfrac{1}{2}\,  \bar{a}^{\mu\nu} \bar{a}_{\mu\nu}+\bar{\varphi}^2 \;. 
  \end{split}
 \ee    
We start by performing a local 
 redefinition to new (primed) fields, implicitly defined as follows 
  \be
  \begin{split}
  h_{\mu\nu} \ &= \ h_{\mu\nu}'+\phi' \eta_{\mu\nu}\;,  \qquad 
  \phi \ = \ \phi'+\tfrac{1}{2}h'\;, \\[1ex] 
  a_{\mu\nu} \ &= \ a_{\mu\nu}'-\varphi' \eta_{\mu\nu}\;, \qquad 
  \varphi \ = \ \varphi'-\tfrac{1}{2}a'\;,\\[1ex] 
  \bar{a}_{\mu\nu} \ &= \ \bar{a}_{\mu\nu}'+\bar{\varphi}' \eta_{\mu\nu}\;, \qquad 
  \bar{\varphi} \ = \ \bar{\varphi}'+\tfrac{1}{2}\bar{a}'\;, 
 \end{split} 
 \ee 
where $h_{\mu\nu}$ is the symmetric (graviton) part of the DFT fluctuation, while fields without 
indices denote the trace parts. The redefinition in the first line brings the theory into Einstein frame. 
Moreover, the redefinition is such that the scalars become gauge singlets. (More precisely, 
for $\varphi, \bar{\varphi}$ this is only a meaningful statement in the tensionless limit, in 
which they are inert under (\ref{gaugeNEW}).) 

Inserting the above field redefinition into the Lagrangian (\ref{evensimplerquadLvmbb1})
and dropping the primes at the end, we arrive at 
 \be\label{CanonicalQuadr}
  \begin{split}
   L \ = \ &-\tfrac{1}{2}h^{\mu\nu} G_{\mu\nu}(h) - \tfrac{1}{12}H^{\mu\nu\rho} H_{\mu\nu\rho}
     \\[1ex]
    &+\tfrac{1}{4} a^{\mu\nu} G_{\mu\nu}(a)+\tfrac{1}{2\alpha'}( a^{\mu\nu} a_{\mu\nu}-\tfrac{1}{2}a^2) \\[1ex]
   &+\tfrac{1}{4} \bar{a}^{\mu\nu} G_{\mu\nu}(\bar a)-\tfrac{1}{2\alpha'}( \bar{a}^{\mu\nu} \bar{a}_{\mu\nu}
   -\tfrac{1}{2}\bar{a}^2) \\[1ex]
   & +\tfrac{1}{4}(D-2)\Big[ \phi\square \phi -\tfrac{1}{2}\varphi\square \varphi 
   -\tfrac{1}{2}\bar{\varphi}\square \bar{\varphi} +\tfrac{2}{\alpha'}\varphi^2 -\tfrac{2}{\alpha'}\bar{\varphi}^2\Big]\;.  
  \end{split}
 \ee  
The first line encodes the kinetic terms for the graviton and the two-form field, 
with the field strength $H_{\mu\nu\rho}=3\partial_{[\mu}b_{\nu\rho]}$ and   
the linearized 
Einstein tensor $G_{\mu\nu}(h)$ defined in (\ref{lin-Einstein-ten}).  

We will now show how the massive fields can be integrated out. 
To illustrate the procedure, let us first recall the simpler case of 
having the massless graviton and a \textit{single} massive spin-2 field, 
\be
  L' \ = \ 
  -\tfrac{1}{2}h^{\mu\nu}G_{\mu\nu}(h) + \tfrac{1}{4}a^{\mu\nu} G_{\mu\nu}(a)
  +\tfrac{1}{2}(a^{\mu\nu} a_{\mu\nu}- a^2)\;, 
 \ee
where we picked the Fierz-Pauli form of the mass term.  
This is  equivalent to a curvature-squared addition for $h_{\mu\nu}$, 
as can be seen as follows: Performing the field redefinition $\bar{h}_{\mu\nu} = h_{\mu\nu}+\frac{1}{\sqrt{2}} a_{\mu\nu}$, 
the Lagrangian becomes 
 \be
  L' \ = \ -\tfrac{1}{2}\bar{h}^{\mu\nu}G_{\mu\nu}(\bar{h}) + \tfrac{1}{\sqrt{2}}a^{\mu\nu} G_{\mu\nu}(\bar{h})
  +\tfrac{1}{2}(a^{\mu\nu} a_{\mu\nu}- a^2)\;,  
 \ee
showing that $a_{\mu\nu}$ is now auxiliary. 
Integrating it out, we obtain a curvature-squared term.\footnote{The specific form of the curvature-squared term is dimension-dependent. For $D=4$ it is equivalent to the square 
of the Weyl tensor, while for $D=3$ (with reversed overall sign) it corresponds to `new massive gravity',  
see sec.~2 of \cite{Bergshoeff:2011ri}.}  

Returning to the full theory, we will apply the same strategy, but since we have two massive 
(ghost-like) fields of each type this will lead to an infinite series of higher-derivative corrections. 
Since the spin-2 and spin-0 
sectors are decoupled, they can be treated separately. 
Focusing first on the scalar part in the last line of the Lagrangian, we perform the 
following redefinition of the dilaton 
 \be
  \phi\,\rightarrow \,\bar{\phi} \ = \ \phi +  \tfrac{1}{\sqrt{2}}(\varphi+\bar{\varphi})\;,  
 \ee
which, after dropping the bar, leads to 
 \be\label{ScalarLang}
  L_{\rm scalar} \ = \  \tfrac{1}{4}(D-2)\Big[ \phi\square \phi -\tfrac{1}{\sqrt{2}}\varphi\square \phi 
   -\tfrac{1}{\sqrt{2}}\bar{\varphi}\square \phi +\varphi\square \bar{\varphi}
   +\tfrac{2}{\alpha'}\varphi^2 -\tfrac{2}{\alpha'}\bar{\varphi}^2\Big]\;.  
 \ee   
We observe that the kinetic terms for $\varphi, \bar{\varphi}$ have cancelled. 
Although there is an off-diagonal term $\varphi\square \bar{\varphi}$, we will show in the following 
that these fields are auxiliary in that they can be eliminated algebraically. 
To this end it is convenient to perform one more change of field basis, introducing 
 \be\label{phipm}
  \varphi^{\pm} \ = \ \varphi \ \pm \ \bar{\varphi}\;, 
 \ee
for which the Lagrangian reads 
 \be\label{newSCalarLa}
   L_{\rm scalar} \ = \  \tfrac{1}{4}(D-2)\Big[ \phi\square \phi -\tfrac{1}{\sqrt{2}}\varphi^+ \square \phi 
   +\tfrac{1}{4}{\varphi}^+\square \varphi^+ -\tfrac{1}{4}\varphi^-\square {\varphi}^{-}
   +\tfrac{2}{\alpha'}\varphi^+\varphi^-\Big]\;.  
  \ee 
The field equations for $\varphi^+$ and ${\varphi}^-$, respectively, can be written as  
 \be\label{phipmeq}
  \begin{split}
   \varphi^- \  &= \ -\tfrac{\alpha'}{4}\square \varphi^+  + 
   \tfrac{\alpha'}{2\sqrt{2}}\square \phi \;, \\[0.5ex]
   \varphi^+ \ &= \ \tfrac{\alpha'}{4}\square \varphi^- \;. 
  \end{split}
 \ee     
Assuming a series expansion of $\varphi^{\pm}$ in positive powers of $\alpha'$, 
these equations can be solved algebraically as follows. 
We observe from the first equation that $\varphi^-$ starts at order 
$\alpha'$, expressed in terms of the dilaton as
$\varphi^- = \tfrac{\alpha'}{2\sqrt{2}}\square \phi +{\cal O}(\alpha^{\prime 2})$, which with the 
second equation implies that $\varphi^+$ vanishes to first order in $\alpha'$. 
More generally, it is easy to see that 
$\varphi^+$ has only contributions for even powers of $\alpha'$, while $\varphi^-$ has only 
contributions for odd powers in $\alpha'$. 
It is then straightforward to give the exact solution for $\varphi^{\pm}$ in terms of the 
dilaton, 
 \be\label{phipm9}
 \begin{split}
  \varphi^+ \  &= \ \sum_{n=1}^{\infty}(\alpha')^{2n}(-1)^{n+1} \frac{1}{\sqrt{2}\,2^{4n-1}}\square^{2n}\phi\ 
   =\ \sqrt{2} \, \frac{1}{ \square^2+ m^4 } \square^2 \phi
   \;, \\[1ex]
   \varphi^- \  &= \ \sum_{n=1}^{\infty}(\alpha')^{2n-1}(-1)^{n+1} \frac{1}{\sqrt{2}\,2^{4n-3}}\square^{2n-1}\phi
   \ =\ \sqrt{2}\,m^2\,  \frac{1}{ \square^2 + m^4} \square \phi
    \;,
 \end{split}
 \ee 
 where $m^2=\tfrac{4}{\alpha'}$.
We can now back-substitute into the Lagrangian (\ref{newSCalarLa}), which is legal because we 
have solved algebraic equations. This computation is simplified by noting that since (\ref{phipm9}) 
satisfy (\ref{phipmeq}) we can rewrite the last term in (\ref{newSCalarLa}): 
 \be
   \tfrac{1}{\alpha'}(\varphi^+\varphi^-+\varphi^+\varphi^-)
  \ = \  \tfrac{1}{4}\varphi^-\square\varphi^--\tfrac{1}{4}\varphi^+\square \varphi^+
  +\tfrac{1}{2\sqrt{2}}\varphi^+\square \phi\;, 
 \ee
where we eliminated in each of the two terms $\varphi^+\varphi^-$ on the left-hand side 
one of the fields according to (\ref{phipmeq}). 
This relation simplifies the Lagrangian by cancelling the `kinetic' terms, while changing the coefficient 
of the term $\varphi^+\square \phi$, leading to\footnote{It is important to emphasize that the resulting Lagrangian 
is only correct when $\varphi^{\pm}$ are expressed in terms of the dilaton according to (\ref{phipm9}). 
Viewed as independent fields such manipulations at the level of the action would be illegal.}  
 \be
  L_{\rm scalar} \ = \ \tfrac{1}{4}(D-2)\Big[\phi\square \phi \ + \ \sum_{n=1}^{\infty} (\alpha')^{2n}(-1)^n
  \frac{1}{2^{4n+1}}\phi\square^{2n+1}\phi\Big]\;. 
 \ee
The infinite series can be rewritten in a closed form as follows  
 \be
  L_{\rm scalar} \ = \ \tfrac{1}{4}(D-2)\Big[\phi\square \phi \ - \tfrac{1}{2}\phi \frac{1}{\square^2+m^4}
  \square^3\phi\Big]\;. 
 \ee
 Thus, the original theory (\ref{ScalarLang}) describing the massless dilaton plus 
two massive (ghost-like) scalars is equivalent to a theory for only the dilaton, but with 
an infinite number of higher-derivative corrections. 

\medskip

We now turn to the problem of integrating out the massive spin-2 fields, which follows precisely the same 
procedure. We start with the spin-2 part of the Lagrangian,  
 \be
  \begin{split}
   L_{\rm spin-2} \ = \ &-\tfrac{1}{2} h^{\mu\nu} G_{\mu\nu}(h)\\[1ex]
   &+\tfrac{1}{4} a^{\mu\nu} G_{\mu\nu}(a)+\tfrac{1}{2\alpha'}( a^{\mu\nu} a_{\mu\nu}-\tfrac{1}{2}a^2) \\[1ex]
   &+\tfrac{1}{4} \bar{a}^{\mu\nu} G_{\mu\nu}(\bar a)-\tfrac{1}{2\alpha'}( \bar{a}^{\mu\nu} \bar{a}_{\mu\nu}
   -\tfrac{1}{2}\bar{a}^2)\;, 
  \end{split}
 \ee  
and perform the field redefinition  
 \be
  \bar{h}_{\mu\nu} \ = \ h_{\mu\nu}+\tfrac{1}{\sqrt{2}}(a_{\mu\nu} + \bar{a}_{\mu\nu})\;. 
 \ee
This field redefinition  breaks $O(D,D)$ covariance because the fields entering here carry different index projections.  As for the scalar case,  
this cancels the kinetic terms for $a$ and $\bar{a}$, and 
we arrive at 
 \be\label{firstORDERspin22} 
  \begin{split}
   L_{\rm spin-2} \ = \  &-\tfrac{1}{2} h^{\mu\nu} G_{\mu\nu}(h)\\[1ex]
   &+\tfrac{1}{\sqrt{2}} a^{\mu\nu} G_{\mu\nu}(h) + \tfrac{1}{\sqrt{2}}\bar{a}^{\mu\nu} G_{\mu\nu}(h)
   -\tfrac{1}{2}a^{\mu\nu}G_{\mu\nu}({\bar a})\\[1ex]
   & +\tfrac{1}{2\alpha'}( a^{\mu\nu} a_{\mu\nu}-\tfrac{1}{2}a^2)
   -\tfrac{1}{2\alpha'}( \bar{a}^{\mu\nu} \bar{a}_{\mu\nu}
   -\tfrac{1}{2}\bar{a}^2)\;, 
  \end{split}
 \ee  
where we dropped the bar on $h$.  
Next, we introduce the field basis, 
 \be 
  a_{\mu\nu}^{\pm} \ = \ a_{\mu\nu} \ \pm \ \bar{a}_{\mu\nu}\;,  
 \ee   
in terms of which the Lagrangian reads  
 \be\label{firstORDERspin222} 
  \begin{split}
   L_{\rm spin-2} \ = \  &-\tfrac{1}{2} h^{\mu\nu} G_{\mu\nu}(h)\\[1ex]
   &+\tfrac{1}{\sqrt{2}} a^{+\mu\nu} G_{\mu\nu}(h) - \tfrac{1}{8}{a}^{+\mu\nu} G_{\mu\nu}(a^+)
   +\tfrac{1}{8}a^{-\mu\nu}G_{\mu\nu}(a^-)\\[1ex]
   & +\tfrac{1}{2\alpha'}( a^{\mu\nu+} a^-_{\mu\nu}-\tfrac{1}{2}a^+ a^-) \;. 
  \end{split}
 \ee  
The field equations for $a^+$ and $a^-$, respectively, then read 
 \be\label{STEPEQQ}
  \begin{split}
   a^{-}_{\mu\nu} - \tfrac{1}{2}a^-\eta_{\mu\nu} \ &= \ \tfrac{\alpha'}{2} G_{\mu\nu}(a^+) 
   -\alpha'\sqrt{2} G_{\mu\nu}(h) \;,
     \\[1ex]
   a^+_{\mu\nu} - \tfrac{1}{2}a^+ \eta_{\mu\nu} \ &= \  -\tfrac{\alpha'}{2}G_{\mu\nu}(a^-)\;. 
  \end{split}
 \ee  
Taking the trace of this equation, we may eliminate the traces  $a^{\pm}$ to obtain  
 \be\label{FINALITeq0}
  \begin{split}
   a^-_{\mu\nu} \ &= \ \tfrac{\alpha'}{2} R_{\mu\nu}(a^+) -\alpha' \sqrt{2} R_{\mu\nu}(h) 
   \;, \\[1ex]
   a^+_{\mu\nu} \ &= \ -\tfrac{\alpha'}{2} R_{\mu\nu}(a^-)\;. 
  \end{split}
 \ee      
As for the scalar case, it is now straightforward to solve these equations iteratively for $a^+$ and $a^-$ 
in powers of $\alpha'$. 
To first order in $\alpha'$ the above equations are solved 
by 
 \be
  a^+_{\mu\nu} \ = \ 0\;, \qquad 
  a^-_{\mu\nu} \ = \ -\alpha'\sqrt{2} R_{\mu\nu}(h)\;, 
 \ee 
while the higher order solutions follow successively by re-inserting into (\ref{FINALITeq0}). 
To this end, one has to note with (\ref{linRicci}) that the Ricci tensor of the Ricci tensor 
takes the form 
 \be
  R_{\mu\nu}(R(h)) \ = \ -\tfrac{1}{2}\square R_{\mu\nu}(h)\;, 
 \ee
where we used the Bianchi identity $\partial^{\mu}R_{\mu\nu}=\frac{1}{2}\partial_{\nu}R$.  
It is straightforward to prove by induction that the solution of (\ref{FINALITeq0}) is 
 \be
  \begin{split}
   a_{\mu\nu}^+ \ &= \ \sum _{n=1}^{\infty}\, (\alpha')^{2n}(-1)^n \frac{\sqrt{2}}{2^{4n-2}}\,\square^{2n-1} 
   R_{\mu\nu}(h)\;,  \\[0.5ex]
    a_{\mu\nu}^- \ &= \ \sum _{n=1}^{\infty}\,(\alpha')^{2n-1}(-1)^{n} \frac{\sqrt{2}}{2^{4n-4}}\,
    \square^{2n-2} R_{\mu\nu}(h)\;.
  \end{split}
 \ee 
Again, since these expressions have been obtained by solving algebraic equations, we can re-insert them into the 
action (\ref{firstORDERspin222}) to obtain 
  \be\label{eq:spin2LagAllOrders}
  L_{\rm spin-2} \ = \ -\tfrac{1}{2} h^{\mu\nu} G_{\mu\nu}(h)  
  \ + \ \sum_{n=1}^{\infty}\,(\alpha')^{2n} (-1)^n \frac{1}{2^{4n-1}}
  \big(R^{\mu\nu}\,\square^{2n-1} R_{\mu\nu}
  -\tfrac{1}{2} R\,\square^{2n-1} R\,\big)\;.
 \ee
As for the scalar sector, we observe that one obtains only corrections of even 
powers in $\alpha'$. 
We also note that this action can be rewritten as 
   \be\label{fullACTION0}
  L_{\rm spin-2} \ = \ -\tfrac{1}{2}\Big(h^{\mu\nu}+4\frac{1}{\square^2+m^4}\square R^{\mu\nu}\Big) G_{\mu\nu}\;, 
 \ee
where $m^2=\tfrac{4}{\alpha'}$. Expanding the geometric series, it is straightforward to
verify the equivalence with the higher-derivative terms in (\ref{eq:spin2LagAllOrders}). 

In total, we have shown that the theory with quadratic Lagrangian (\ref{CanonicalQuadr}) is equivalent 
to the theory for the massless graviton, Kalb-Ramond field and dilaton, but with an infinite 
number of higher-derivative corrections:   \be\label{FINALNSNSACtion}
  \begin{split}
   L \ = \ &-\tfrac{1}{2}\Big(h^{\mu\nu}+4\frac{1}{\square^2+m^4}\square R^{\mu\nu}\Big) G_{\mu\nu} - \tfrac{1}{12}H^{\mu\nu\rho} H_{\mu\nu\rho}
   \\[1ex]
   &+\ \tfrac{1}{4}(D-2)\Big[\phi\square \phi \ - \ \tfrac{1}{2}\phi \frac{1}{\square^2+m^4}\square^3\phi\Big]\;.  
  \end{split}
 \ee  
As the higher-derivative terms are proportional to the Einstein tensor $G_{\mu\nu}(h)$  or to $\square\phi$, it is clear that to any finite order in $\alpha'$
we may remove these corrections by a local field redefinition of the metric and the dilaton.  
In order to remove the complete infinite series would 
require a nonlocal redefinition. 
Such redefinitions are illegal when it comes to proving the equivalence of two theories, 
and therefore our result is not in conflict with the presence of extra physical and massive modes 
in the quadratic theory. 
One can also integrate out the massive fields without sacrificing manifest T-duality \cite{unUnpublished}, finding  a Lagrangian that is physically equivalent to the one in equation (\ref{FINALNSNSACtion}).

\section*{Acknowledgments}

We  thank Nathan Berkovits, Diego Marques, Lionel Mason,  Ashoke Sen and Warren Siegel 
for useful discussions. 
O.H.~is supported by a DFG Heisenberg Fellowship 
of the German Science Foundation (DFG).  The work of U.N.~and B.Z.~is supported by the U.S. Department of Energy under grant Contract Number de-sc0012567.

\begin{appendix}

\sectiono{Degrees of freedom}

The purpose of this appendix is to determine the physical degrees of freedom propagated by the HSZ theory and the massive deformation of DFT.  
We will abandon manifest $O\bkt{D,D}$ invariance 
by taking the derivatives $D$ and $\bar D$ to be
partial derivatives, using indices as $\mu,\nu, \ldots$, and $\square = \partial^2$.
We start by determining the spectrum of the two-derivative part of the HSZ theory. Then we compare it with the spectrum of the full HSZ theory and show that no extra degrees of freedom appear upon adding higher derivative terms. Finally, we  consider the spectrum for the massive deformation of DFT given in section 4 and show that it does not propagate any ghost-like degree of freedom.

\subsection{Degrees of freedom of two-derivative HSZ theory}\label{App:DOF-two-der}

Consider the two derivative quadratic Lagrangian given in equation (\ref{eq:L20L22}).
The part of the Lagrangian involving $e_{ij}$ and $\phi$ is trivial to analyze and it describes massless graviton, dilaton and two-form field. We thus focus on the part of the Lagrangian involving the $a$-field. After putting in explicit factors of $\alpha'$, we have:
\be\label{evensimplerquadLvm}\
    L \ = 
    - \tfrac{1}{8}\,  a^{\mu\nu} \, \square 
    \, a_{\mu\nu}  -\tfrac{1}{4} (\partial_{\mu}a^{\mu\nu})^2
     \ + \tfrac{1}{2\alpha'}\,  a^{\mu\nu} a_{\mu\nu}\;.
    \ee   
  We rescale the field as $a \to 2a$ to get the canonical normalization for the kinetic term and also define $m^2 = \frac{4}{\alpha'}$. After coupling to a source $J_{\mu\nu}$ the Lagrangian becomes:
  \be\label{evensimplerquadLvmsp}\
    L \ = 
    - \tfrac{1}{2}\,  a^{\mu\nu} \, (\square  
    - m^2)  \, a_{\mu\nu} 
     + a^{\mu\nu} \partial_\mu \partial^\rho a_{\rho\nu}
     +  a^{\mu\nu} J_{\mu\nu}\;.
    \ee   
The equations of motion in momentum space take the following form
\be
\label{eom-mom-space1}
   (p^2+  m^2)  \, a_{\mu\nu}   -p_\mu (p\cdot a)_\nu     -p_\nu (p\cdot a)_\mu  
\ = \  -  J_{\mu\nu}\,.    \ee 
Using the equations of motion in the Lagrangian it takes the following form in the momentum space:
\be
\label{Lsimplified-vm0}
L \ = \ \tfrac{1}{2}   \, J^{\mu\nu} \bkt{-p}a_{\mu\nu}\bkt{p}   \,. 
\ee
Let us introduce the notation $\bkt{pap} \equiv p_\mu a^{\mu\nu} p_\nu $ and $\bkt{p\cdot a}_\mu = p^\nu a_{\nu\mu}$. Contracting the above equation with $p^\mu p^\nu$ and solving for $\bkt{pap}$, we get:
 \be
 (pap) \ = \ {(pJp)\over p^2 - m^2}  \,  \,. 
\ee
Contracting  equation (\ref{eom-mom-space1}) with $p^\nu$ and using the expression for $\bkt{pap}$, we can solve for $\bkt{p a}_\mu$:

\be
\label{first-contraction-followup} 
(p\cdot a)_\mu \ = \  {1\over m^2} \Bigl(     \frac{p_\mu}{p^2-m^2}  (pJp) 
- (p\cdot J)_\mu \Bigr) \,.
\ee
Using these expressions for $\bkt{pap}$ and $\bkt{pa}_{\mu}$ in (\ref{eom-mom-space1}) we can solve for $a_{\mu\nu}$ and obtain:
\be
  a_{\mu\nu} \ = \  - \, {1\over p^2+ m^2} \Bigl(  {J}_{\mu\nu} +{1\over m^2} \Bigl( p_\mu (p\cdot J)_\nu  
+ p_\nu (p\cdot J)_\mu  \Bigr)  \Bigr)    \, + {2\, p_\mu p_\nu \over m^2}  {(pJp)\over( p^2 - m^2)(p^2+ m^2)}   \,. 
\ee
Decomposing the last term into partial fractions, we get
\be
  a_{\mu\nu} \ = \  - \, {1\over p^2+ m^2} \widetilde {J}_{\mu\nu}    \, +{1\over( p^2 - m^2)} {p_\mu p_\nu (pJp)\over m^4}     \,, 
\ee
where
\be\label{eq:JtildDef}  \widetilde{J}_{\mu\nu}  \ \equiv \ {J}_{\mu\nu} +{1\over m^2} \bigl( p_\mu (p\cdot J)_\nu  
+ p_\nu (p\cdot J)_\mu  \bigr)  + {p_\mu p_\nu (pJp)\over m^4} \;. 
\ee
Back in (\ref{Lsimplified-vm0}) the Lagrangian becomes
\be
L \ = \ -\tfrac{1}{2} J^{\mu\nu}\bkt{-p} {1\over p^2+m^2}  \widetilde J_{\mu\nu}\bkt{p} 
+ \tfrac{1}{2m^4} (pJp) (-p)  {1\over p^2 - m^2}  {(pJp)\bkt{p}} \,. 
\ee
The nature of the degrees of freedom is determined by the residues at the poles. 
At the pole $p^2+m^2=0$, it is easy to see that $\tilde{J}^{\mu\nu}$ is transverse, i.e., $p_\mu \widetilde{J}^{\mu\nu}\ =0$. Using this, we can write the Lagrangian as follows:
\be
L \ = \ -\tfrac{1}{2} \widetilde{J}^{\mu\nu}\bkt{-p} {1\over p^2+m^2}  \widetilde J_{\mu\nu}\bkt{p} 
+ \tfrac{1}{2m^4} (pJp) (-p)  {1\over p^2 - m^2}  {(pJp)\bkt{p}}\; .
\ee
The first term implies that we are propagating a ghostly (overall minus sign) massive
spin two mode (the traceless part of $\tilde{J}_{\mu\nu}$) and a ghostly, massive  scalar (the trace
of $\tilde{J}_{\mu\nu}$), both with mass squared equal to $m^2$.  The second term shows a proper tachyonic  scalar with mass squared given by $-m^2$.

The analysis of the sector involving $\bar{a}_{\mu\nu}$ can be done similarly. Note that the kinetic terms for $\bar{a}_{\mu\nu}$ and $a_{\mu\nu}$ have the same sign but their mass terms have opposite signs. Hence, the $\bar{a}_{\mu\nu}$ sector describes a ghostly tachyonic spin-2, a ghostly tachyonic scalar and a healthy massive scalar.

\subsection{Degrees of freedom of full quadratic HSZ theory}\label{a2app}

Here we consider the full quadratic theory as given in the Lagrangian (\ref{evensimplerquadL}). We see that the three sectors, $\bkt{e_{\mu\nu},\phi}$,  $\bkt{a_{\mu\nu}, \varphi}$ and $\bkt{\bar{a}_{\mu\nu},\bar{\varphi}}$ are completely decoupled. 
The sector $\bkt{e_{\mu\nu},\phi}$ is well known and describes a massless graviton, dilaton, and $b$-field. Let us just focus on the $\bkt{a_{\mu\nu},\varphi}$ sector of the Lagrangian given by:
\be\label{evensimplerquadLvm9}\
    L \ = 
    - \tfrac{1}{8}\,  a^{\mu\nu} \, \square 
    \, a_{\mu\nu}  -\tfrac{1}{4} (\partial_{\mu}a^{\mu\nu})^2
    -\tfrac{1}{2}\,\partial_\mu a^{\mu\nu} \partial_\nu\varphi  \ + \tfrac{1}{2\alpha'}\,  a^{\mu\nu} a_{\mu\nu} 
    \, +\, \tfrac{1}{2\alpha'}\varphi  \square  
    \varphi  
   -\varphi^2 \;,
    \ee   
    where we have put explicit factors of $\alpha'$.
    We rescale the field $a_{\mu\nu}\to 2a_{\mu\nu}$ to get a canonical kinetic term and define $m^2\equiv \tfrac{4}{\alpha'}$. After coupling to sources $J_{\mu\nu}$ and $K$ the Lagrangian takes the following form
  \be\label{evensimplerquadLvmsp9}\
    L \ = 
    - \tfrac{1}{2}\,  a^{\mu\nu} \, (\square
    - m^2)  \, a_{\mu\nu} 
     + a^{\mu\nu} \partial_\mu \partial^\rho a_{\rho\nu}
    + \, a^{\mu\nu} \partial_\mu\partial_\nu\varphi +\, \tfrac{1}{2}\varphi(\square 
     -\tfrac{1}{2} m^2) \varphi   +  a^{\mu\nu} J_{\mu\nu}  + \varphi K \;. 
    \ee  
The equations of motion in momentum space are given by:
\be
\label{eom-mom-space}
  \begin{split}
  (p^2+  m^2)  \, a_{\mu\nu}   -p_\mu (p\cdot a)_\nu     -p_\nu (p\cdot a)_\mu  
  - p_\mu p_\nu \varphi \ = \ & -  J_{\mu\nu}\,, \\[1.0ex]
  (p^2 +\tfrac{1}{2} m^2) \varphi  + (pap) \ = \ & \  K \,. 
  \end{split}
  \ee 
Using the equations of motion in the Lagrangian it takes the following form in the momentum space:
\be
\label{Lsimplified-vm}
L \ = \ \tfrac{1}{2}   \, J^{\mu\nu}\bkt{-p} a_{\mu\nu}\bkt{p}   \, + \, \tfrac{1}{2} \, K\bkt{-p} \varphi\bkt{p}  \,. 
\ee
 Contracting the first equation in (\ref{eom-mom-space}) with $p^\mu$ we get
\be
\label{first-contraction} 
m^2\,  (p\cdot a)_\nu     -p_\nu \bigl( (pap) + p^2 \varphi \bigr)  \ = \  -  (p\cdot J)_{\nu}\;. 
\ee
Contracting this with $p^\mu$ and solving for $(pap)$ we obtain:
\be
 (pap) \ = \ {(pJp)\over p^2 - m^2}  \, - \, {p^4 \varphi\over p^2-m^2} \,, 
\ee
where we notice tachyonic poles (that will disappear later).  Using this in the  equation of motion for $\varphi$ (second one in (\ref{eom-mom-space})), we can solve for $\varphi$ and obtain:
\be
\label{varphi-sol}
\phantom{\Biggl(} \ 
\varphi  \ = \ {2\over m^2} \cdot  {(pJp)\over p^2 + m^2}  \, - \, {2\over m^2}\cdot  
{p^2-m^2\over p^2+m^2} \,  K  \,.  \ \ 
\ee
We now reconsider the first contraction (\ref{first-contraction}) to find
\be
\label{first-contraction-followup} 
(p\cdot a)_\mu\ = \  {1\over m^2} \Bigl(     p_\mu  \bigl( (pap) + p^2 \varphi \bigr) 
- (p\cdot J)_ \mu\Bigr)\;. 
\ee
Using this and the expression for $\varphi$ in the equation of motion for $a_{\mu\nu}$ we can solve for $a_{\mu\nu}$ in terms of sources and get:
\be
a_{\mu\nu} \ = \  - \,  \frac{1}{p^2+m^2}\widetilde{J}_{\mu\nu}     + \,  { p_\mu p_\nu \over m^2 \bkt{p^2+m^2}} \bkt{ {\bkt{pJp} \over m^2}+ 2 K} \,, 
\ee
where
$ \widetilde{J}_{\mu\nu} $ is the transverse part of $J_{\mu\nu}$ as defined in equation (\ref{eq:JtildDef}).
Back  in (\ref{Lsimplified-vm}) the Lagrangian becomes
\be
\begin{split}
L \ = \ &-\tfrac{1}{2} J^{\mu\nu}\bkt{-p} {1\over p^2+m^2} \widetilde J_{\mu\nu}\bkt{p}  - \ K\bkt{-p} \frac{p^2-m^2}{m^2\bkt{p^2+m^2}} K\bkt{p} \\[1ex]
& +\bkt{pJp}\bkt{-p}\frac{1}{m^2\bkt{p^2+m^2}} K\bkt{p}+ K (-p) \frac{1}{m^2\bkt{p^2+m^2}}\bkt{pJp}\bkt{p}\;. 
\end{split}
\ee

We now have to look at the pole $p^2+m^2 = 0$. Using the fact that $\tilde{J}_{\mu\nu}$ is transverse at the pole, the Lagrangian can be written in the following form at the pole:
\be
\begin{split}
L \ = \ &-\tfrac{1}{2} \widetilde{J}^{\mu\nu}\bkt{-p} {1\over p^2+m^2} \widetilde J_{\mu\nu}\bkt{p}  +\tfrac{1}{2} \ \Bigl( 2K\ + {\bkt{pJp}\over m^2}\Bigr) \bkt{-p} \frac{1}{p^2+m^2} 
\Bigl( 2K\ + {\bkt{pJp}\over m^2}\Bigr)\bkt{p} \;. 
\end{split}
\ee
The first term tells us that we are propagating a ghostly (overall minus sign) massive 
spin-2 mode (the traceless part of $\widetilde{J}_{\mu\nu}$) and a ghostly, massive scalar (the trace
of $\widetilde{J}_{\mu\nu}$).  The second term shows a proper massive scalar.

The analysis of the $\bkt{\bar{a}_{\mu\nu},\bar{\varphi}}$ sector can be done similarly. Since the mass terms of the two sectors have opposite signs and the kinetic terms have the same sign, the $\bkt{\bar{a}_{\mu\nu},\bar{\varphi}}$ sector propagates a ghostly tachyonic spin-2, a ghostly tachyonic scalar and a proper tachyonic scalar.
 If we compare this spectrum with that of the two-derivative theory we see that 
 the {\em full} spectrum remains unchanged.

\subsection{Degrees of freedom of massive DFT}\label{App:mDFT}

We start with the Lagrangian for the massive DFT as given in equation (\ref{massiveDFT}). We scale the fields as $e_{\mu\nu}\to \sqrt{2}e_{\mu\nu}$ and $ \phi\to \frac{1}{\sqrt{2}}\phi$ to get canonical normalization for the kinetic terms. By using $e_{\mu\nu}=h_{\mu\nu}+b_{\mu\nu}$, the Lagrangian for the massive DFT can be written as:
\be 
L_{\text{mDFT}}= L_{h,\phi}\ + L_b\;,
\ee
where
\be 
\begin{split}
L_{h,\phi} \ &= \ \tfrac{1}{2} h^{\mu\nu} \square  
h_{\mu\nu} + \bkt{\p_\mu h^{\mu\nu}}^2  
          +h^{\mu\nu}\p_\mu   \p_\nu\phi -   \tfrac{1}{2}\phi\, \square 
          \phi -\tfrac{1}{2}M^2(h^{\mu\nu} h_{\mu\nu}-  \phi^2)\;,  \label{eq:LmDFTCan} \\[1ex]
L_{b}\ &=\ \tfrac{1}{2} b^{\mu\nu} \square  
b_{\mu\nu} +\bkt{\p_\mu b^{\mu\nu}} ^2\ -\tfrac{1}{2}M^2 b^{\mu\nu}b_{\mu\nu}\;.
\end{split}
\ee
The Lagrangian $L_b$ is well known to describe a massive two-form field and will
not be discussed further.  
In order to make it clear that the mass terms in $L_{h, \phi}$ are special,
we modify one of the coefficients by introducing a parameter $\gamma$.  We will
indeed find that the value $\gamma =1$ is selected by the condition that we have
no ghosts in the spectrum.  We thus take, henceforth, 
\be 
\begin{split}
L_{h,\phi} \ &= \ \tfrac{1}{2} h^{\mu\nu} \square 
h_{\mu\nu} + \bkt{\p_\mu h^{\mu\nu}}^2  
          +h^{\mu\nu}\p_\mu   \p_\nu\phi -   \tfrac{1}{2}\phi\, \square
                    \phi -\tfrac{1}{2}M^2(h^{\mu\nu} h_{\mu\nu}- \gamma \phi^2)\;.  \label{eq:LmDFTCan2} 
          \end{split}
\ee
After coupling to sources $J_{\mu\nu}$ and $K$ 
 for $h_{\mu\nu}$ and $\phi$, we have:
\be 
L_{h,\phi} \ = \ \tfrac{1}{2} h^{\mu\nu} \square 
h_{\mu\nu} + \bkt{\p_\mu h^{\mu\nu}}^2  
          +h^{\mu\nu}\p_\mu \p_\nu\phi -\tfrac{1}{2}\phi\, \square            \phi -\tfrac{1}{2}M^2(h^{\mu\nu} h_{\mu\nu}-  \gamma\phi^2)\ + J^{\mu\nu}h_{\mu\nu}\ + K \phi\;. \label{eq:LmDFThphiS}
\ee
In momentum space, the equations of motion take the following form:
\be 
\begin{split}
\label{eq:eomhphib}
J_{\mu\nu}-\bkt{p^2+M^2} h_{\mu\nu}+2p_{(\mu}p_{\rho}h^\rho{}_{\nu)}- p_{\mu}p_{\nu}\phi \ &= \ 0\;,\ \\[1ex]
K+\bkt{p^2+\gamma M^2}\phi\ -p_{\mu}p_{\nu}h^{\mu\nu} \ &= \ 0\; .
\end{split}
\ee
Using these the Lagrangian takes the following form in the momentum space:
\be 
L_{h,\phi}\ =\ \tfrac{1}{2}J^{\mu\nu}\bkt{-p}h_{\mu\nu}\bkt{p} + \tfrac{1}{2}K\bkt{-p} \phi\bkt{p}\; .
\ee
Contracting the top equation of motion with $p^\mu p^\nu$ we get:
\be 
\bkt{pJp}\ + \bkt{p^2-M^2}\bkt{php}- p^4 \phi\ =\ 0\;.
\ee
The above equation and the equation of motion for $\phi$ can now be used to eliminate $\phi$ and $\bkt{php}$ in favor of sources:
\be 
\begin{split}
\phi\ &=\ 
-\frac{\bkt{pJp}}{M^2 A}\  - \frac{p^2-M^2}{M^2 A} K \\
\bkt{php}\ &=\ -\frac{p^4}{M^2 A} K\ -\frac{p^2+\gamma M^2}{M^2 A} \bkt{pJp}\;, \label{eq:phpPhi2}
\end{split}
\ee
where $A$ is given by
\be 
A\ =\ p^2\bkt{\gamma-1}\ -\gamma M^2\,  .
\ee
Contracting the equation of motion for $h_{\mu\nu}$ with $p^\nu$ we get:
\be 
\bkt{pJ}_\mu\ + p_{\mu}\bkt{php-p^2 \phi}\ =\ M^2\bkt{ph}_\mu.
\ee
Using eqns. (\ref{eq:phpPhi2}) yields 
\be 
\bkt{ph}_\mu= \frac{1}{M^2} \bkt{pJ}_\mu\ -\frac{p_\mu}{M^2 A}\bkt{p^2 K + \gamma\bkt{pJp}}.\label{eq:(ph)i2}
\ee
Finally, using eqns. (\ref{eq:(ph)i2}) and (\ref{eq:phpPhi2}) in the equation of motion for $h_{\mu\nu}$ we obtain 
\be 
h_{\mu\nu}\ = \widetilde{J}_{\mu\nu}\ \frac{1}{p^2+M^2} \ - \frac{p_\mu p_\nu}{M^4A} \bkt{M^2 K\ + \bkt{\gamma-1} \bkt{pJp}}\;,
\ee
where $\tilde{J}_{\mu\nu}$ is defined by
\be 
\widetilde{J}_{\mu\nu}\ = \ J_{\mu\nu}\ +\frac{2}{M^2} p_{(\mu}\bkt{ph}_{\nu)}\ + \frac{p_\mu p_\nu }{M^4}\bkt{pJp}\;.
\ee
It is easy to see that on the mass-shell  $p^2=-M^2$, the tensor $\tilde{J}_{\mu\nu}$ is transverse,
\be 
\label{transverse-vm}
p^\mu \widetilde{J}_{\mu\nu}\ =\ 0\;  \quad ( p^2=-M^2) \, .
\ee 
This will be useful below.
Inserting these expressions back into $L_{h,\phi}$,
 we get:
\be  \label{lbhexp}
L_{h,\phi}\, =\,  \tfrac{1}{2}  
 J_{\mu\nu}\bkt{-p} \frac{1}{p^2+M^2} \widetilde{J}^{\mu\nu}\bkt{p} \ 
 -  \tfrac{1}{2} \ \frac{\bkt{pJp}^2\bkt{\gamma-1} +2M^2\bkt{pJp} K + M^2K^2 \bkt{p^2-M^2}}{M^4\bkt{\gamma-1}\bkt{p^2\ +\frac{\gamma M^2}{1-\gamma}}}\;. 
\ee
For the case of interest, $\gamma=1$,
 the second term above is completely regular and
 we need only focus on the first term.  Using the transversality condition
 (\ref{transverse-vm}) we can rewrite $L_{h,\phi}$ as
\be 
L_{h,\phi}\ =\ \widetilde{J}_{\mu\nu}\bkt{-p}\ \frac{1}{p^2+M^2} \widetilde{J}^{\mu\nu}\bkt{p} 
+ \ldots \; ,
\ee
near $p^2= -M^2$ and where the dots indicate terms that are regular.
At the mass-shell we can choose $p = (M, \vec{0})$ and thus the transversality
condition implies that $\widetilde J_{0\mu} = \widetilde J_{\mu0} = 0$.   The only non vanishing components
of $\widetilde J_{\mu\nu}$ are those where both indices represent spatial directions.
We are thus propagating $(D-1)D/2$ positive-norm degrees of freedom, 
associated with a symmetric $(D-1) \times (D-1)$ matrix.
The trace-less part corresponds to the massive spin-2 and the trace corresponds to the massive scalar. 

For the case $\gamma \not=1$ the above degrees of freedom are
still present but we now have more, due to the pole in the second term of (\ref{lbhexp}).
This time the mass-shell is  $p^2\ =\ -\frac{\gamma M^2}{1-\gamma}$
and we go to a Lorentz frame where $p^0\ =\sqrt{\frac{\gamma M^2}{1-\gamma}}$. 
Near the pole we now find
\be   
L_{h,\phi}|_{\text{second pole}}\ =\ - \tfrac{1}{2} \ 
\frac{(K-\gamma J_{00})(-p)\ (K-\gamma J_{00})(p)}{\bkt{\gamma-1}^2 \bkt{p^2\ +\frac{\gamma M^2}{1-\gamma}}}  \ + \ldots  \,,
\ee
making it manifest that for $\gamma \neq  1$ we propagate an additional ghostly massive scalar.
We conclude that the model constructed in section (4.1) describes massive graviton, dilaton and $b$-field and does not propagate any extra undesired degrees of freedom.

\end{appendix}

\end{document}